\documentclass[usenatbib,usegraphicx]{mn2e}
\usepackage{times}

\title[Data boundary fitting]{Data boundary fitting using a generalised 
least-squares method}
\author[N. Cardiel]{N. Cardiel$^{1}$\thanks{E-mail:
ncl@astrax.fis.ucm.es}\\
$^{1}$Departamento de Astrof\'{\i}sica y CC.\ de la Atm\'{o}sfera, 
Facultad de Ciencias F\'{\i}sicas, 
Ciudad Universitaria s/n, E28040--Madrid, Spain}

\begin{document}

\date{Accepted \ldots Received \ldots; in original form \ldots}

\pagerange{\pageref{firstpage}--\pageref{lastpage}} \pubyear{2009}

\maketitle

\label{firstpage}

%%%%%%%%%%%%%%%%%%%%%%%%%%%%%%%%%%%%%%%%%%%%%%%%%%%%%%%%%%%%%%%%%%%%%%%%%%%%%%%
\begin{abstract}
In many astronomical problems one often needs to determine the upper and/or
lower boundary of a given data set. An automatic and objective approach
consists in fitting the data using a generalised least-squares method, where
the function to be minimized is defined to handle \emph{asymmetrically} the
data at both sides of the boundary. In order to minimise the cost
function, a numerical approach, based on the popular {\sc downhill} simplex
method, is employed. The procedure is valid for any numerically computable
function.  Simple polynomials provide good boundaries in common situations. For
data exhibiting a complex behaviour, the use of \emph{adaptive splines} gives
excellent results. Since the described method is sensitive to extreme data
points, the simultaneous introduction of error weighting and the flexibility of
allowing some points to fall outside of the fitted frontier, supplies the
parameters that help to tune the boundary fitting depending on the nature of
the considered problem. Two simple examples are presented, namely the
estimation of spectra pseudo-continuum and the segregation of scattered data
into ranges. The normalisation of the data ranges prior to the fitting
computation typically reduces both the numerical errors and the number of
iterations required during the iterative minimisation procedure.
\end{abstract}

%%%%%%%%%%%%%%%%%%%%%%%%%%%%%%%%%%%%%%%%%%%%%%%%%%%%%%%%%%%%%%%%%%%%%%%%%%%%%%%

%%%%%%%%%%%%%%%%%%%%%%%%%%%%%%%%%%%%%%%%%%%%%%%%%%%%%%%%%%%%%%%%%%%%%%%%%%%%%%%
\section{Introduction}
\label{section:introduction}

Astronomers usually face, in their daily work, the need of determining the
boundary of some data sets. Common examples are the computation of frontiers
segregating regions in diagrams (e.g.\ colour--colour plots), or the estimation
of reasonable pseudo-continua of spectra. Using for illustration the latter
example, several strategies are initially feasible in order to get an
analytical determination of that boundary. One can, for example, fit a simple
polynomial to the general trend of the considered spectrum, masking previously
disturbing spectroscopic features, such as important emission lines or deep
absorption characteristics. Since this fit \emph{traverses} the data, it must be
shifted upwards a reasonable amount in order to be placed on top of the
spectrum. However, since there is no reason to expect the pseudo-continuum
following exactly the same functional form as the polynomial fitted through the
spectrum, that shift does not necessarily provides the expected answer. As an
alternative, one can also force the polynomial to pass over some special
data points, which are selected to guide (actually to force) the fit through
the apparent upper envelope of the spectrum. With this last method the result
can be too much dependent on the subjectively selected points. In any case,
the technique requires the additional effort of determining those special
points.

With the aim of obtaining an objective determination of the boundaries, an
automatic approach, based on a generalisation of the popular least-squares
method, is presented in this work. Section~\ref{section:the_method} describes
the procedure in the general case. As an example, the boundary fitting using
simple polynomials is included in this section. Considering that these simple
polynomials are not always flexible enough,
Section~\ref{section:adaptive_splines} presents the use of \emph{adaptive
splines}, a variation of the typical fit to splines that allows the
determination of a boundary that smoothly adapts to the data in an iterative
way.  Section~\ref{section:application} shows two practical uses of this
technique: the computation of spectra pseudo-continuum and the determination of
data ranges. Since the scatter of the data due to the presence of data
uncertainties tends to bias the boundary determinations,
Section~\ref{section:uncertainties} analyses the problem and presents a
modification of the method that allows to confront this situation.  Finally
Section~\ref{section:conclusions} summarises the main conclusions. In addition,
Appendix~\ref{appendix:constraints} discusses the inclusion of constraints in
the fits, whilst Appendix~\ref{appendix:normalization} describes how the
normalisation of the data ranges prior to the data fitting can help to reduce
the impact of numerical errors in some circumstances.

The method described in this work has been implemented into the program {\tt
BoundFit}, a FORTRAN code written by the author and available (under the GNU
General Public License\footnote{See license details at {\tt http://fsf.org}},
version 3) at the following URL\\
{\small\tt http://www.ucm.es/info/Astrof/software/boundfit}\\
All the fits presented in this paper have been computed with this program.

%%%%%%%%%%%%%%%%%%%%%%%%%%%%%%%%%%%%%%%%%%%%%%%%%%%%%%%%%%%%%%%%%%%%%%%%%%%%%%%
\section{A generalised least-squares method}
\label{section:the_method}

%------------------------------------------------------------------------------
\subsection{Introducing the asymmetry}
\label{subsection:asymmetry}

\begin{figure}
\includegraphics[angle=-90,width=\columnwidth]{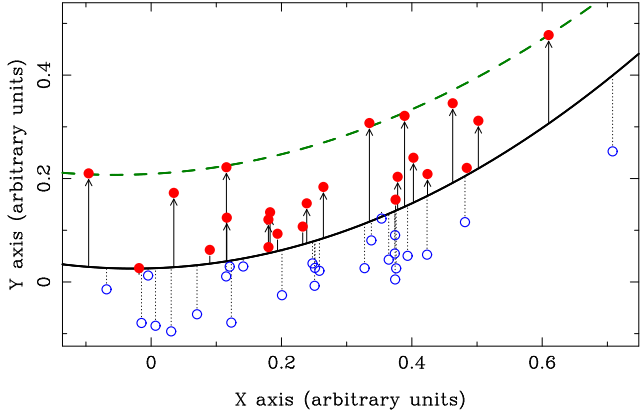}
\caption{Graphical illustration of the asymmetrical weighting scheme described
in Section~\ref{subsection:asymmetry} for the determination of the upper
boundary of a particular data set. In this example a second-order polynomial is
employed. The continuous thick line is the traditional (symmetric) ordinary
least-squares fit for the whole set of data points, which is used as an
initial guess for the boundary determination. The filled red circles are data
points above that fit (i.e.\ outside), whereas the open blue
circles are found below such frontier (inside). Filled circles receive the
extra weighting factor parametrized by the asymmetry coefficient~$\xi$
introduced in Eq.~(\ref{equation:asymmetry}). Since this parameter is chosen to
be $\xi>>1$, the minimisation process shifts the initial fit towards the upper
region. By iterating the procedure, the final boundary fit, shown as the green
dashed line, is obtained. The same method, but exchanging symbols weights,
could be employed to determine the lower boundary limit (not shown).}
\label{figure:cartoon}
\end{figure}

The basic idea behind the method that follows is to introduce, in the fitting
procedure, an asymmetric role for the data at both sides of a given fit, so the
points located outside relative to that fit pull stronger toward
themselves than the points at the opposite side. This idea is graphically
illustrated in Fig.~\ref{figure:cartoon}. As it is going to be shown, the
problem is numerically treatable. In order to use the data asymmetrically,
it is necessary to start with some initial guess fit, that in practice
can be obtained employing the traditional least-squares method (with a
symmetric data treatment). Once this initial fit is available, it is
straightforward to continue using the data asymmetrically and, in an iterative
process, determine the sought boundary.

Let's consider the case of a two-dimensional data set consisting in $N$ points
of coordinates $(x_i,y_i)$, where $x_i$ is an independent
variable, and $y_i$ a dependent variable, which value has an associated and
known uncertainty $\sigma_i$. An ordinary error-weighted least-squares fit is
obtained by minimising the \emph{cost function} $f$ (also called
\emph{objective function} in the literature concerning optimisation
strategies), defined as
\begin{equation}
f(a_0,a_1,\ldots,a_p) =
\sum_{i=1}^{N} \left( \frac{y(x_i)-y_i}{\sigma_i} \right)^2,
\label{equation:ols}
\end{equation}
where $y(x_i)$ is the \emph{fitted function} evaluated at \mbox{$x=x_i$}, and
$a_0,a_1,\ldots,a_p$ are the unknown $(p+1)$ parameters that define such
function. Actually, one should write the fitted function as
\mbox{$y(a_0,a_1,\ldots,a_p;x)$}.

In order to introduce the asymmetric weighting scheme, the cost function can be
generalised introducing some new coefficients,
\begin{equation}
f(a_0,a_1,\ldots,a_p) =
\sum_{i=1}^{N} w_i | y(x_i)-y_i |^\alpha,
\label{equation:gls}
\end{equation}
where $\alpha$ is now a variable exponent (\mbox{$\alpha=2$} in normal 
least squares).  For that reason the distance between the fitted function $y(x_i)$
and the dependent variable $y_i$ is considered in absolute value. The new
overall weighting factors $w_i$ are defined differently depending on whether
one is fitting the upper or the lower boundary. More precisely
\begin{equation}
w_i \equiv \left\{
\begin{array}{ll}
\begin{array}{@{}c@{}}\textrm{upper}\\ \textrm{boundary}\end{array} &
    \left\{ \begin{array}{ll}
                  1/\sigma_i^\beta & \textrm{for}\;\; y(x_i) \ge y_i\\
                \xi/\sigma_i^\beta & \textrm{for}\;\; y(x_i) < y_i
     \end{array} \right. \\
                        & \\
\begin{array}{@{}c@{}}\textrm{lower}\\ \textrm{boundary}\end{array} &
     \left\{ \begin{array}{ll}
                \xi/\sigma_i^\beta & \textrm{for}\;\; y(x_i) > y_i\\
                  1/\sigma_i^\beta & \textrm{for}\;\; y(x_i) \le y_i
     \end{array} \right.
\end{array}
\right.
\label{equation:asymmetry}
\end{equation}
being $\beta$ the exponent that determines how error weighting is incorporated
into the fit (\mbox{$\beta=0$} to ignore errors, \mbox{$\beta=2$} in normal
error-weighted least squares), and $\xi$ is defined as an \emph{asymmetry
coefficient}.  Obviously, for \mbox{$\alpha=\beta=2$} and \mbox{$\xi=1$},
Eq.~(\ref{equation:gls}) simplifies to Eq.~(\ref{equation:ols}). As it is going
to be shown later, the asymmetry coefficient must satisfy \mbox{$\xi >>1$} for
the method to provide the required boundary fit.

Leaving apart the particular weighting effect of the data
uncertainties~$\sigma_i$, the net outcome of introducing the factors~$w_i$ is
that the points that are classified as being outside from a given frontier
simply have a higher weight that the points located at the inner side (see
Fig.~\ref{figure:cartoon}), and this difference scales with the particular
value of the asymmetry coefficient~$\xi$.

Thus, the boundary fitting problem reduces to finding the \mbox{$(p+1)$}
parameters \mbox{$a_0,a_1,\ldots,a_p$} that minimise Eq.~(\ref{equation:gls}),
subject to the weighting scheme defined in Eq.~(\ref{equation:asymmetry}).
In the next sections several examples are provided, in which the functional
form of $y(x)$ is considered to be simple polynomials and splines.

%------------------------------------------------------------------------------
\subsection{Relevant issues}

The method just described is, as defined, very sensitive to extreme data
points. This fact, that at first sight may be seen as a serious problem, it is
not necessarily so. For example, one may be interested in constraining the
scatter exhibited by some measurements due to the presence error sources.
In this case a good option would be to derive the upper and lower
frontiers that surround the data, and in this scenario there is no need to
employ an error-weighting scheme (i.e.\ \mbox{$\beta=0$} would be the
appropriate choice). On the other hand, there are situations in which the data
sample contains some points that have larger uncertainties than others, and one
wants those points to be ignored during the boundary estimation. Under
this circumstance the role of the $\beta$ parameter in
Eq.~(\ref{equation:asymmetry}) is important. Given the relevance of all these
issues concerning the impact of data uncertainties in the boundary computation,
this topic is intentionally delayed until Section~\ref{section:uncertainties}.
At this point it is better to keep the problem in a more simplified
version, which facilitates the examination of the basic properties of the
proposed fitting procedure.

An interesting generalisation of the boundary fitting method described above
consists in the incorporation of additional constraints during the minimisation
procedure, like forcing the fit to pass through some predefined fixed points,
or imposing the derivatives to have some useful values at particular
points. A discussion about this topic has been included in
Appendix~\ref{appendix:constraints}.

Another issue of great relevance is the appearance of numerical errors during
the minimisation procedure. The use of data sets exhibiting values with
different orders of magnitude, or with a very high number of data points, can
be responsible for preventing numerical methods to provide the expected
answers. In some cases a simple solution to these problems consists in
normalising the data ranges prior to the numerical minimisation. A detailed
description of this approach is presented in
Appendix~\ref{appendix:normalization}.

%------------------------------------------------------------------------------
\subsection{Example: boundary fitting to simple polynomials}
\label{subsection:simple_polynomial}

Returning to Eq.~(\ref{equation:gls}), let's consider now the particular case
in which the functional form of the fitted boundary $y(x)$ is assumed to be a
simple polynomial of degree $p$, i.e.
\begin{equation}
y(x)=a_0+a_1 x+a_2 x^2+\ldots+a_p x^p.
\end{equation}
In this case, the function to be minimized, \mbox{$f(a_0,a_1,\ldots,a_p)$}, is
also a simple function of the \mbox{$(p+1)$} coefficients. In ordinary least
squares one simply takes the partial derivatives of the cost function $f$ with
respect to each of these coefficients, obtaining a set of \mbox{$(p+1)$}
equations with \mbox{$(p+1)$} unknowns, which can be easily solved, as far as
the number of independent points $N$ is large enough, i.e.\ $N\ge p+1$.

However, considering the special definition of the weighting coefficients $w_i$
given in Eq.~(\ref{equation:asymmetry}), it is clear that in the general
case an analytical solution cannot be derived without any kind of iterative
approach, since during the computation of the considered boundary (either
upper or lower), the classification of a particular data point as being inside
or outside relative to a given fit explicitly depends on the function $y(x)$
that one is trying to derive. Fortunately numerical minimisation
procedures can provide the sought answer in an easy way. For this purpose, the
{\sc downhill} simplex method \citep{downhill} is an excellent option. This
numerical procedure performs the minimisation of a function in a
multi-dimensional space. For this method to be applied, an initial guess for
the solution must be available. This initial solution, together with a
characteristic length-scale for each parameter to be fitted, is employed to
define a simplex (i.e., a multi-dimensional analogue of a triangle) in the
solution space. The algorithm works using only function evaluations (i.e.\ not
requiring the computation of derivatives), and in each iteration the method
improves the previously computed solution by modifying one of the vertices of
the simplex. The simplex adapts itself to the local landscape, and contracts on
to the final minimum. The numerical procedure is halted once a pre-fixed
numerical precision in the sought coefficients is reached, or when the number
of iterations exceeds a pre-defined maximum value $N_{\mbox{\scriptsize
maxiter}}$. A well-known implementation of the {\sc downhill} simplex method is
provided by \citet{numrec}\footnote{Since the Numerical Recipes license is too
restrictive (the routines cannot be distributed as source), the implementation
of {\sc downhill} included in the program {\tt BoundFit} is a personal version
created by the author to avoid any legal issue, and as such it is distributed
under the GNU General Public License, version3.}. For the particular case of
minimising Eq.~(\ref{equation:gls}) while fitting a simple polynomial, a
reasonable guess for the initial solution is supplied by the coefficients of an
ordinary least-squares fit to a simple polynomial derived by minimising
Eq.~(\ref{equation:ols}).

It is important to highlight that whatever the numerical method employed to
perform the numerical minimisation, the considered cost function will probably
exhibit a parameter-space landscape with many peaks and valleys. The finding of
a solution is never a guarantee of having found the right answer, unless one
has the resources to employ brute force to perform a really exhaustive search
at sufficiently fine sampling of the cost function to find the global minimum.
In situations where this problem can be serious, more robust methods, like
those provided by genetic algorithms, must be considered \citep[see
e.g.][]{genetic}. Fortunately, for the particular problems treated in this
paper, the simpler {\sc downhill} method is a good alternative, considering
that the ordinary least-squares method will likely give a good initial guess
for the expected solution in most of the cases.

\begin{figure}
\includegraphics[angle=0,width=\columnwidth]{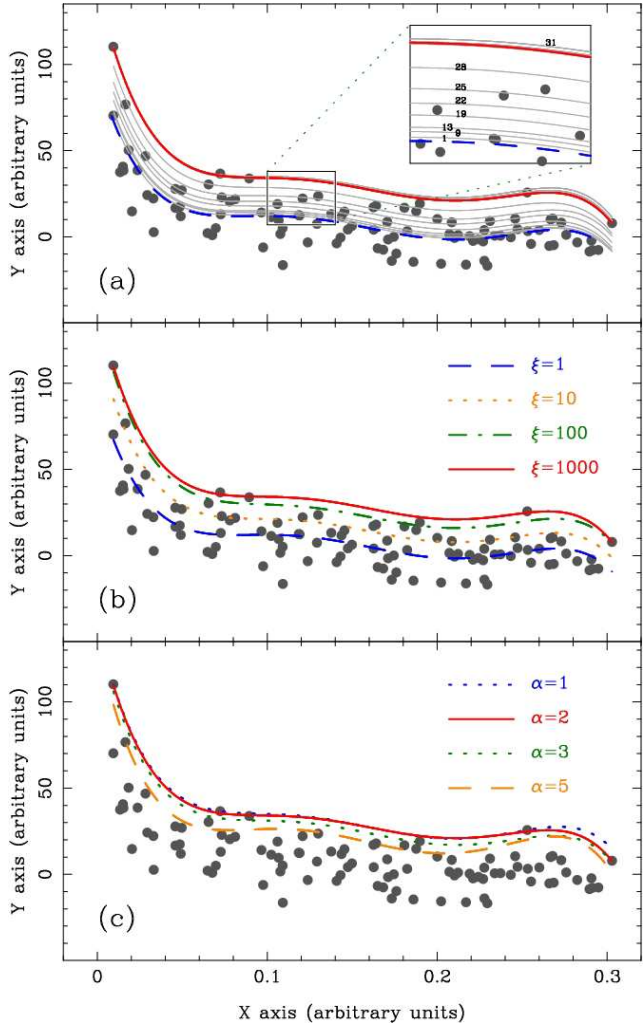}
\caption{\emph{Panel~(a)}: Example of upper boundary fitting using a 5th order
polynomial. The initial data set correspond to 100 points randomly drawn from
the function \mbox{$y(x)=1/x$}, assuming the uncertainty \mbox{$\sigma=10$} for
all the points in the $y$-axis. The dashed blue line is the ordinary 
least-squares fit to that data, used as the initial guess for the numerical
determination of the boundary. Since all the points have the same uncertainty,
there is no need for an error-weighted procedure. For that reason
\mbox{$\beta=0$} has been used in Eq.~(\ref{equation:asymmetry}). In addition
\mbox{$\alpha=2$} and an asymmetry coefficient \mbox{$\xi=1000$} were employed.
The grey lines indicate the boundary fits obtained for $N_{\mbox{\scriptsize
maxiter}}$ in the range from 5 to 2000~iterations, at arbitrary steps. The
inset displays a zoomed plot region where some particular values of
$N_{\mbox{\scriptsize maxiter}}$ are annotated over the corresponding fits. The
continuous red line is the final boundary determination obtained using
\mbox{$N_{\mbox{\scriptsize maxiter}}=2000$}. \emph{Panel~(b)}: Effect of
employing different asymmetry coefficients $\xi$ for the upper boundary fit
shown in panel~(a). In the four cases the same maximum number of iterations
\mbox{($N_{\mbox{\scriptsize maxiter}}=2000$)} has been employed, with
\mbox{$\alpha=2$}. \emph{Panel~(c)}: Effect of using different values of the
power $\alpha$, with \mbox{$N_{\mbox{\scriptsize maxiter}}=2000$}
and~\mbox{$\xi=1000$}. See discussion in
Section~\ref{subsection:simple_polynomial}.}
\label{figure:oneoverx_pol}
\end{figure}

For illustration, Fig.~\ref{figure:oneoverx_pol}a displays an example of upper
boundary fitting to a given data set, using a simple 5th order polynomial. As
initial guess for the numerical minimisation, the ordinary least-squares fit
for the data (shown with a dashed blue line) has been employed. The grey lines
represent the corresponding boundary fits obtained using the {\sc downhill}
method previously described. Each line corresponds to a pre-defined maximum
number of iterations $N_{\mbox{\scriptsize maxiter}}$ in {\sc downhill}, as
labelled over the lines in the plot inset. In this particular example the
fitting procedure has been carried out without weighting with errors (i.e.,
assuming \mbox{$\beta=0$}), and using a power \mbox{$\alpha=2$} and an
asymmetry coefficient \mbox{$\xi=1000$}. It is clear that after a few
iterations the intermediate fits move upwards from the initial guess (dashed
blue line), until reaching the location marked with $N_{\mbox{\scriptsize
maxiter}}=31$. Beyond this number of iterations, the fits move downwards
slightly, rapidly converging into the final fit displayed with the continuous
red line. Fig.~\ref{figure:oneoverx_pol}b displays the effect of modifying the
asymmetry coefficient $\xi$. The ordinary least-squares fit corresponds to
\mbox{$\xi=1$} (dashed blue line). The asymmetric fits are obtained for
$\xi>1$. The figure illustrates how for \mbox{$\xi=10$} and~100 the resulting
upper boundaries do still leave points in the \emph{wrong} side of the
boundary. Only when \mbox{$\xi=1000$} (continuous red line) is the boundary fit
appropriate. Thus, a proper boundary fitting requires the asymmetry coefficient
to be large enough to compensate for the pulling effect of the points that are
in the inner side of the boundary.  On the other hand,
Fig.~\ref{figure:oneoverx_pol}c shows the impact of changing the power $\alpha$
in Eq.~(\ref{equation:gls}). For the lowest value, \mbox{$\alpha=1$} (dotted
blue line), the fit is practically identical to the one obtained with
\mbox{$\alpha=2$} (continuous red line).  For the largest values,
\mbox{$\alpha=3$ or~5} (dotted green and dashed orange lines), the boundaries
are below the expected location, leaving some points outside (above) the fits.
In these last cases the power $\alpha$ is too high and, for that reason, the
distance from the boundary to the more distant points in the inner side have a
too high effect in the cost function given by Eq.~(\ref{equation:gls}).

\begin{figure}
\includegraphics[angle=0,width=\columnwidth]{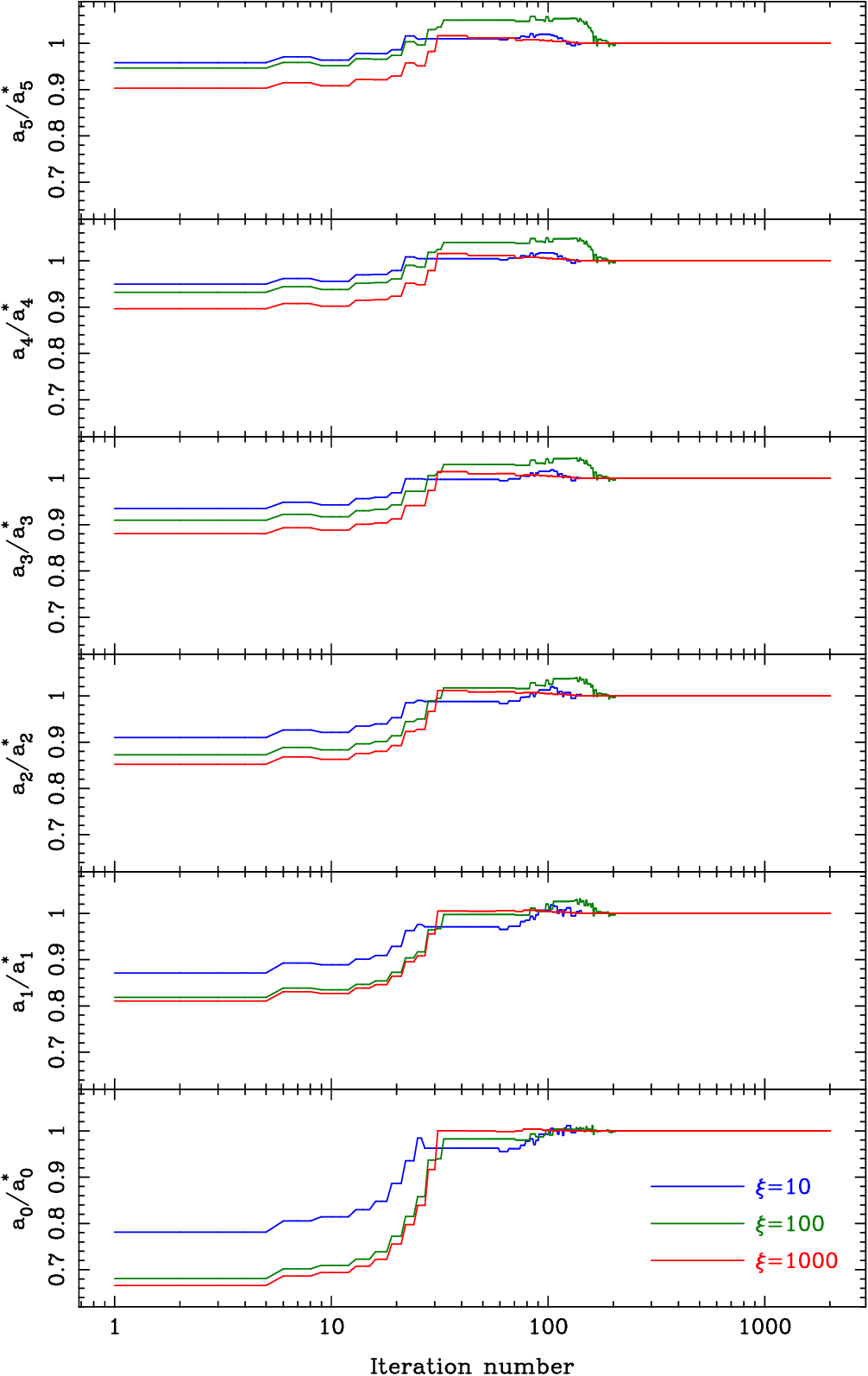}
\caption{Variation in the fitted coefficients, as a function of the number of
iterations, for the upper boundary fit (5th order polynomial
\mbox{$y(x)=\sum_{i=0}^5 a_i\;x^i$}) shown in
Fig.~\ref{figure:oneoverx_pol}a. Each panel represents the coefficient value
at a given iteration ($a_i$, with \mbox{$i=0,\ldots,5$}, from bottom to top)
divided by $a_i^{*}$, the final value derived after \mbox{$N_{\mbox{\scriptsize
maxiter}}=2000$} iterations. The same $y$-axis range is employed in all the
plots. Red lines correspond to an asymmetry coefficient \mbox{$\xi=1000$},
whereas the blue and green grey lines indicate the coefficients obtained with
\mbox{$\xi=10$} and \mbox{$\xi=100$}, respectively (in all the cases
\mbox{$\alpha=2$} and \mbox{$\beta=0$} have been employed). Note that the plot
$x$-scale is in logarithmic units.}
\label{figure:coeff_pol}
\end{figure}

Another important aspect to take into account when using a numerical method is
the convergence of the fitted coefficients. Fig.~\ref{figure:coeff_pol}
displays, for the same example just described in
Fig.~\ref{figure:oneoverx_pol}b, the values of the 6~fitted polynomial
coefficients as a function of the maximum number of iterations allowed. The
figure includes the results for \mbox{$\xi=10$}, 100 and~1000 (using
\mbox{$\alpha=2$} and \mbox{$\beta=0$} in the three cases). In overall, the
convergence is reached faster when $\xi=1000$.
Fig.~\ref{figure:oneoverx_pol}a already showed that for this particular value
of the asymmetry coefficient a quite reasonable fit is already achieved when
\mbox{$N_{\mbox{\scriptsize maxiter}}$=31}. Beyond this maximum number of
iterations the coefficients only change slightly, until they definitely settle
around \mbox{$N_{\mbox{\scriptsize maxiter}}\sim 140$}.

Although simple polynomials can be excellent functional forms for a boundary
determination (as shown in the previous example), when the data to be fitted
exhibit rapidly changing values, a single polynomial is not always able to
reproduce the observed trend. A powerful alternative in these situations
consists in the use of splines. The next section presents an improved method
that using classic cubic splines, but introducing additional degrees of
freedom, offers a much larger flexibility for boundary fitting.

%%%%%%%%%%%%%%%%%%%%%%%%%%%%%%%%%%%%%%%%%%%%%%%%%%%%%%%%%%%%%%%%%%%%%%%%%%%%%%%
\section{Adaptive Splines}
\label{section:adaptive_splines}

%------------------------------------------------------------------------------
\subsection{Using splines with adaptable knot location}

Splines are commonly employed for interpolation and modelling of arbitrary
functions. Many times they are preferred to simple polynomials due to their
flexibility. A spline is a piecewise polynomial function that is locally very
simple, typically third-order polynomials (the so called cubic splines). These
local polynomials are forced to pass through a prefixed number of points,
$N_{\mbox{\scriptsize knots}}$, which we will refer as knots. In this way, the
functional form of a fit to splines can be expressed as
\begin{equation}
\begin{array}{@{}r@{\;}l}
y(x)  = & s_3(k) [x-x_{\mbox{\scriptsize knot}}(k)]^3 +
          s_2(k) [x-x_{\mbox{\scriptsize knot}}(k)]^2 + \nonumber \\
      + & s_1(k) [x-x_{\mbox{\scriptsize knot}}(k)] + s_0(k),
\end{array}
\end{equation}
where ($x_{\mbox{\scriptsize knot}}(k),y_{\mbox{\scriptsize knot}}(k)$) are the
$(x,y)$ coordinates of the $k^{\mbox{\scriptsize th}}$~knot, and $s_0(k)$,
$s_1(k)$, $s_2(k)$, and $s_3(k)$ are the corresponding spline coefficients for
\mbox{$x\in[x_{\mbox{\scriptsize knot}}(k),x_{\mbox{\scriptsize knot}}(k+1)]$},
with \mbox{$k=1,\ldots,N_{\mbox{\scriptsize knots}}-1$}. These coefficients are
easily computable by imposing the set of splines to define a continuous
function and that, in addition, not only the function, but also the first and
second derivatives match at the knots (two additional conditions are required;
typically they are provided by assuming the second derivatives at the two
endpoints to be zero, leading to what are normally referred as \emph{natural
splines}). The computation of splines is widely described in the literature
(see e.g. \citealt{splines}).

The final result of a fit to splines will strongly depend on both, the number
and the precise location of the knots. With the aim of having more flexibility
in the fits, \citet{cardiel99} explored the possibility of setting the location
of the knots as free parameters, in order to determine the optimal coordinates
of these knots that improve the overall fit of the data. The solution to the
problem can be derived numerically using any minimisation algorithm, as the
{\sc downhill} simplex method previously described. In this way the set of
splines smoothly adapts to the data. The same approach can be
applied to the data boundary fitting, using as functional form for the function
$y(x)$ in Eq.~(\ref{equation:gls}) the \emph{adaptive splines} just described.
It is important to highlight that in this case the optimal boundary fit
requires not only to find the appropriate coefficients of the splines, but also
the optimal location of the knots.

%------------------------------------------------------------------------------
\subsection{The fitting procedure}

In order to carry out the double optimisation process (for the coefficients and
the knots location) required to compute a
boundary fit using adaptive splines, the following steps can be followed:

\begin{enumerate}

\item \emph{Fix the initial number of knots to be employed},
$N_{\mbox{\scriptsize knots}}$. Using a large value provides more flexibility,
although the number of parameters to be determined logically scales with this
number, and the numerical optimisation demands a larger computational effort.

\item \emph{Obtain an initial solution with fixed knot locations}. For this
purpose it is sufficient, for example, to start by dividing the full $x$-range
to be fitted by \mbox{$(N_{\mbox{\scriptsize knots}}-1)$}. This leads to a
regular distribution of equidistant knots. The initial fit is then derived by
minimising the cost function given in Eq.~(\ref{equation:gls}), leaving as free
parameters the $y$-coordinates of all the knots simultaneously, while keeping
fixed the corresponding $x$-coordinates. This numerical fit also requires a
preliminary guess solution, than can be easily obtained through
\mbox{$(N_{\mbox{\scriptsize knots}}-1)$} independent ordinary least-squares
fit of the data placed between each consecutive pair of knots, using for this
purpose simple polynomials of degree~1 or~2. In this guess solution the
$y$-coordinate for each knot is then evaluated as the average value for the two
neighbouring preliminary polynomial fits (only one for the knots at the borders
of the $x$-range). Obviously, if there is additional information concerning a
more suitable knot arrangement than the equidistant pattern, it must be used to
start the process with an even better initial solution which will facilitate a
faster convergence to the final solution.

\item \emph{Refine the fit}. Once some initial spline coefficients have been
determined, the fit is refined by setting as free parameters the location of
all the “inner” knots, both in the $x$- and $y$-directions. The outer knots
(the first and last in the ordered sequence) are only allowed to be refined in
the $y$-axis direction with the aim of preserving the initial $x$-range
coverage. The simultaneous minimisation of the $x$ and $y$ coordinates of all
the knots at once will imply finding the minimum of a multidimensional function
with too many variables. This is normally something very difficult, with no
guarantee of a fast convergence. The problem reveals to be treatable just by
solving for the optimised coordinates of every single knot separately. In
practice, a \emph{refinement} can be defined as the process of refining the
location of all the $N_{\mbox{\scriptsize knots}}$ knots, one at a time, where
the order in which a given knot is optimised is randomly determined. Each knot
optimisation requires, in turn, a value for the maximum number of iterations
allowed $N_{\mbox{\scriptsize maxiter}}$.  Thus, at the end of every single
refinement process all the knots have been refined once. An extra penalisation
can be introduced in the cost function with the idea of avoiding that knots
exchange their order in the list of ordered sequence of knots. This inclusion
typically implies that, if $N_{\mbox{\scriptsize knots}}$ is large, several
knots end up colliding and having the same coordinates.The whole process
can be repeated by indicating the total number of refinement processes,
$N_{\mbox{\scriptsize refine}}$.

\item \emph{Optimise the number of knots}. If after $N_{\mbox{\scriptsize
refine}}$ refinement processes several knots have collided and exhibit the same
coordinates, this is an evidence that $N_{\mbox{\scriptsize knots}}$ was
probably too large. In this case, those colliding knots can be merged and the
effective number of knots be accordingly reduced.  If, on the contrary, the
knots being used do not collide, it is interesting to check whether a higher
$N_{\mbox{\scriptsize knots}}$ can be employed.  With the new
$N_{\mbox{\scriptsize knots}}$, step~(iii) is repeated again. 

\end{enumerate}

Although at first sight it may seem excessive to use a large number of knots
when some of them are going to end up colliding, these collisions will
typically take place at optimised locations for the considered fit. As far as
the minimisation algorithm is able to handle such large $N_{\mbox{\scriptsize
knots}}$, it is not such a bad idea to start using an overestimated number
and merge the colliding knots as the refinement processes take place.

The fitting algorithm can be halted once a satisfactory fit is found at the end
of step~(iii). By satisfactory one can accept a fit which coefficients do not
significantly change by increasing neither $N_{\mbox{\scriptsize refine}}$ nor
$N_{\mbox{\scriptsize maxiter}}$, and in which there are no colliding knots.

%------------------------------------------------------------------------------
\subsection{Example: boundary fitting to adaptive splines and comparison with
simple polynomials}
\label{subsection:example_splines}

\begin{figure}
\includegraphics[angle=0,width=\columnwidth]{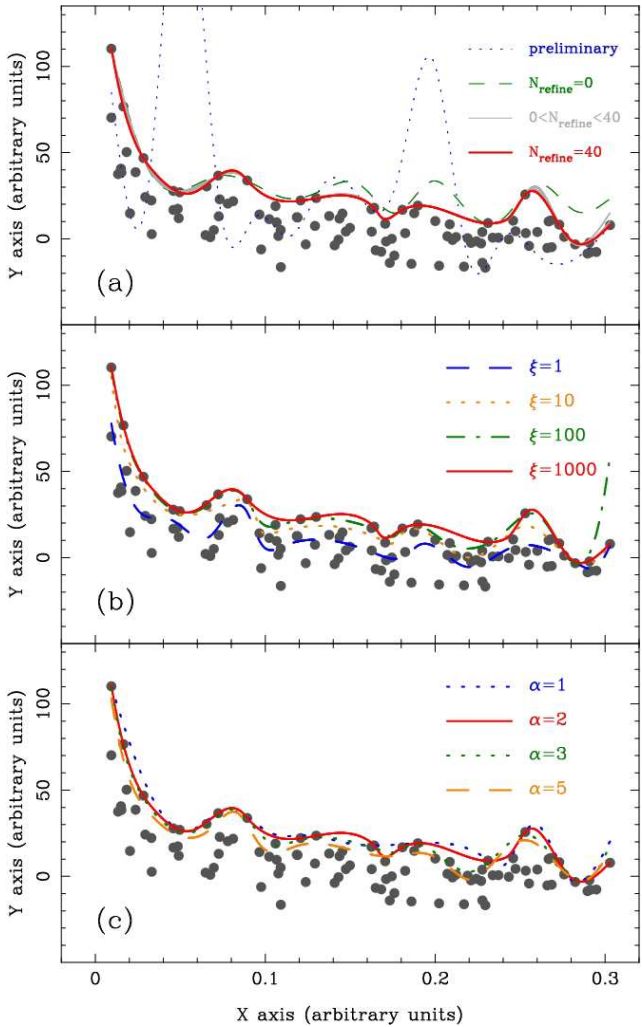}
\caption{Example of the use of adaptive splines to compute the upper boundary
of the same sample data displayed in Fig.~\ref{figure:oneoverx_pol}. In this
case \mbox{$N_{\mbox{\scriptsize knots}}=15$} has been employed.
\emph{Panel~(a)}: the preliminary fit (dotted blue line) shows the initial
guess determined from \mbox{$(N_{\mbox{\scriptsize knots}}-1)$} independent
ordinary least-squares fit of the data, as explained in
Section~\ref{subsection:example_splines}. By imposing
\mbox{$N_{\mbox{\scriptsize maxiter}}=1000$} the fit improves, although in
most cases the effective $N_{\mbox{\scriptsize maxiter}}$ is much lower since
the algorithm computes spline coefficients that have converged before the
number of iterations reaches that maximum value. The dashed green line shows
the first fit obtained with still the knots at their initial equidistant
locations. Successive refinements (light grey) allow the knots to change
their positions, which leads to the final boundary determination (continuous
red line, corresponding to \mbox{$N_{\mbox{\scriptsize refine}}=40$}). In all
these fits \mbox{$\xi=1000$}, \mbox{$\alpha=2$} and \mbox{$\beta=0$} have been
employed. \emph{Panel~(b)}: Effect of using different asymmetry coefficients
$\xi$ for the upper boundary fit shown in the previous panel. In the four cases
\mbox{$N_{\mbox{\scriptsize maxiter}}=1000$}, \mbox{$N_{\mbox{\scriptsize
refine}}=40$}, \mbox{$\alpha=2$} and~\mbox{$\beta=0$} were used.
\emph{Panel~(c)}: Effect of employing different values of the power $\alpha$,
with \mbox{$\xi=1000$}, \mbox{$N_{\mbox{\scriptsize refine}}=40$}
and~\mbox{$\beta=0$}. See discussion in
Section~\ref{subsection:example_splines}.}
\label{figure:oneoverx_spl}
\end{figure}

\begin{figure}
\includegraphics[angle=0,width=\columnwidth]{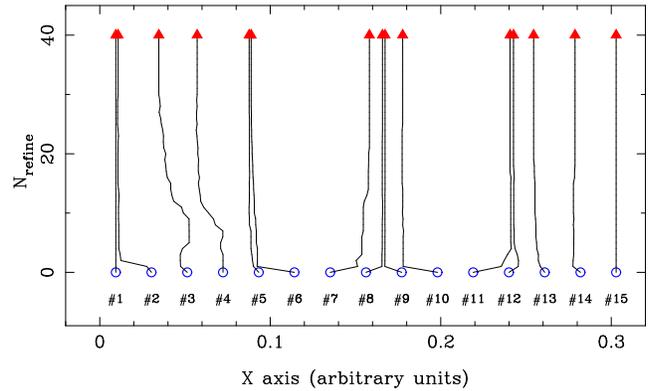}
\caption{Variation in the location of the knots corresponding to the upper
boundary fitting to adaptive splines displayed in
Fig.~\ref{figure:oneoverx_spl}a. Before introducing any refinement
\mbox{($N_{\mbox{\scriptsize refine}}=0$)}, the 15 knots were regularly placed,
as shown with the open blue circles. In each refinement process the inner knots
are allowed to modify its location, one at a time. The first and last knots are
fixed in order to preserve the fitted \mbox{$x$-range}. The final knot
locations after \mbox{$N_{\mbox{\scriptsize refine}}=40$} are shown with the
filled red triangles.}
\label{figure:knots}
\end{figure}

To illustrate the flexibility of adaptive splines,
Fig.~\ref{figure:oneoverx_spl}a displays the corresponding upper boundary fit
employing the same example data displayed in Fig.~\ref{figure:oneoverx_pol},
for the case \mbox{$N_{\mbox{\scriptsize knots}}=15$}. The preliminary fit
(shown as a dotted blue line) was computed by placing the
$N_{\mbox{\scriptsize knots}}$ equidistantly spread in the $x$-axis range
exhibited by the data, and performing \mbox{$(N_{\mbox{\scriptsize knots}}-1)$}
independent ordinary least-squares fit of the data placed between each
consecutive pair of knots, using 2nd order polynomials, as explained in
step~(ii). Although unavoidably this preliminary fit is far from the final
result (due to the fact that this is just the merging of several independent
ordinary fits \emph{through} data exhibiting large scatter and that the
$x$-range between adjacent knots is not large), after $N_{\mbox{\scriptsize
maxiter}}$ iterations without any refinement (i.e., without modifying the
initial equidistant knot pattern) the algorithm provides the fit shown as the
dashed green line. The light grey lines display the resulting fits obtained by
allowing the knot locations to vary, and after 40 refinements one gets the
boundary fit represented by the continuous red line. Since the knot location
has a large influence in the quality of the boundary determination, very high
values for $N_{\mbox{\scriptsize maxiter}}$ are not required (typically values
for the number of iterations needed to obtain refined knot coordinates
are~$\sim 100$). Analogously to what was done with the simple polynomial fit,
in Fig.~\ref{figure:oneoverx_spl}b and~\ref{figure:oneoverx_spl}c the effects
of varying the asymmetry coefficient $\xi$ and the power $\alpha$ are also
examined. In the case of $\xi$, it is again clear that the highest value
\mbox{($\xi=1000$)} leads to a tighter fit. Concerning the power $\alpha$,
the best result is obtained when distances are considered quadratically, i.e.
\mbox{$\alpha=2$}. For the largest values, \mbox{$\alpha=3$} and~5, the
resulting boundaries leave points above the fits. The case \mbox{$\alpha=1$} is
not very different to the quadratic fit, although in some regions (e.g.
\mbox{$x\in[0.01,0.04]$}) the boundary is probably too high. In addition,
Fig.~\ref{figure:knots} displays the variation in the location of the knots as
$N_{\mbox{\scriptsize refine}}$ increases, for the final fit displayed in
Fig.~\ref{figure:oneoverx_spl}a. The initial equidistant pattern (open blue
circles; corresponding to \mbox{$N_{\mbox{\scriptsize refine}}=0$}) is modified
as each individual knot is allowed to change its coordinates. It is clear that
some of the knots approximate and could be, in principle, merged into single
knots, revealing that the initial number of knots was overestimated.

\begin{figure}
\includegraphics[angle=-90,width=\columnwidth]{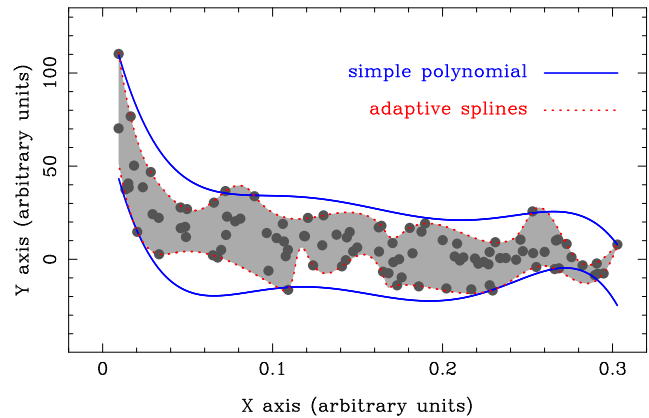}
\caption{Comparison between different functional forms for the boundary
fitting. The sample data set corresponds to the same values employed in
Figs.~\ref{figure:oneoverx_pol} and~\ref{figure:oneoverx_spl}. The boundaries
have been determined using simple polynomials of 5th degree (continuous blue
lines) and adaptive splines (dotted red lines; \mbox{$N_{\mbox{\scriptsize
knots}}=15$} and \mbox{$N_{\mbox{\scriptsize refine}}= 40$}), following the
steps given in Sections~\ref{subsection:simple_polynomial}
and~\ref{subsection:example_splines}, respectively. The shaded area is simply
the diagram region comprised between both adaptive splines boundaries. As
expected, adaptive splines are more flexible, providing tighter boundaries than
simple polynomials.}
\label{figure:example}
\end{figure}

Finally Fig.~\ref{figure:example} presents, for the same sample data employed
in Figs.~\ref{figure:oneoverx_pol} and~\ref{figure:oneoverx_spl}, the
comparison between the boundary fits to simple polynomials (continuous blue
lines) and to adaptive splines (dotted red lines). The shaded area corresponds
to the diagram region comprised between the two adaptive splines boundaries. In
this figure both the upper and the lower boundary limits, computed as described
previously, are represented. It is clear from this graphical comparison that
the larger number of degrees of freedom introduced with adaptive splines allows
a much tighter boundary determination. The answer to the immediate
question of which fit (simple polynomials or splines) is more appropriate
will obviously depend on the nature of the considered problem.

%%%%%%%%%%%%%%%%%%%%%%%%%%%%%%%%%%%%%%%%%%%%%%%%%%%%%%%%%%%%%%%%%%%%%%%%%%%%%%%
\section{Practical applications}
\label{section:application}

%------------------------------------------------------------------------------
\subsection{Estimation of spectra pseudo-continuum}
\label{section:application_pseudo}

As mention in Section~\ref{section:introduction}, a typical situation in which
the computation of a boundary can be useful is in the estimation of spectra
pseudo-continuum. 
The strengths of spectral features have been measured in different ways so
far. However, although with slight differences among them, most authors have
employed line-strength indices with definitions close to the classical
expression for an equivalent width
\begin{equation}
\mbox{EW(\AA)}=\displaystyle\int_{\mbox{line}} (1-S(\lambda)/C(\lambda)
\mbox{d}\lambda,
\end{equation}
where $S(\lambda)$ is the observed spectrum and $C(\lambda)$ is the local
continuum, usually obtained by interpolation of $S(\lambda)$ between two
adjacent spectral regions \citep[e.g.][]{faber73,faber77,whitford83}.  In
practice, as pointed out by \citet{geisler84} \citep[see also][]{rich88}, at
low and intermediate spectral resolution the local continuum is unavoidably
lost, and a pseudo-continuum is measured instead of a true continuum.
The upper boundary fitting, either by using simple
polynomials or adaptive splines, constitutes an excellent option for the
estimation of that pseudo-continuum.
To illustrate this statement, several examples are presented and discussed in
this section. In all these examples, the boundary fits have been computed
ignoring data uncertainties, i.e., assuming \mbox{$\beta=0$} in
Eq.~(\ref{equation:asymmetry}). The impact of errors is this type of
application is discussed later, in Section~\ref{section:uncertainties}.

\begin{figure*}
\includegraphics[angle=-90,width=\textwidth]{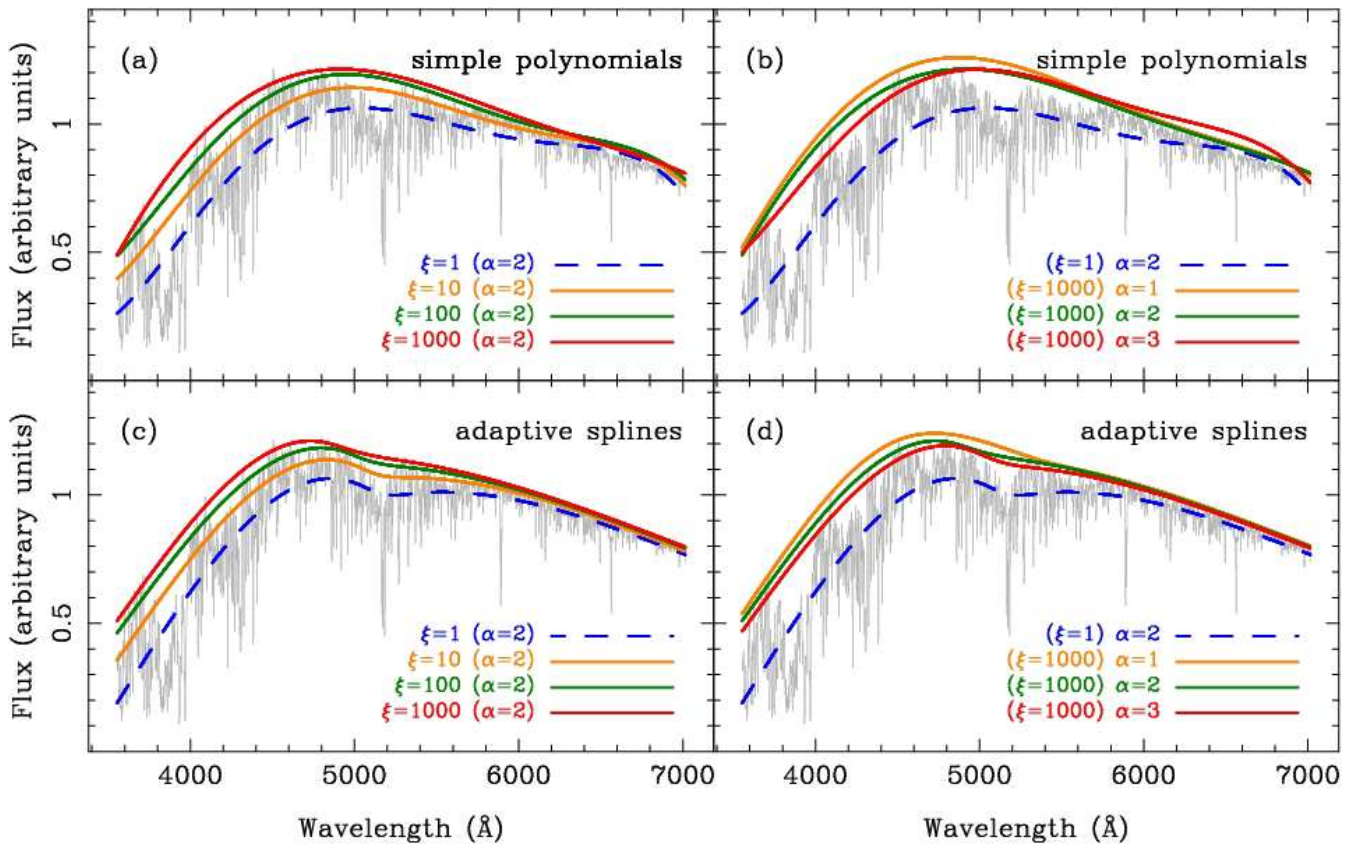}
\caption{Examples of pseudo-continuum fits derived using upper boundaries with
different tunable parameters. Panels~(a) and~(b) correspond to simple 5th order
polynomials, whereas adaptive splines have been employed in panels~(c) and
~(d). The stellar spectrum corresponds to the K0V star HD003651 belonging to
the MILES library \citep{miles}. In the four panels the dashed blue line
indicates the ordinary least-squares fit of the data. See discussion in
Section~\ref{section:application_pseudo}.}
\label{figure:pol_spl}
\end{figure*}

Fig.~\ref{figure:pol_spl} displays upper boundary fits for the particular
stellar spectrum of HD003651 belonging to the MILES\footnote{See {\tt
http://www.ucm.es/info/Astrof/miles/}} library \citep{miles}. The results using
simple polynomials and adaptive splines with different tunable parameters are
shown. Panels~\ref{figure:pol_spl}a and~\ref{figure:pol_spl}b show the results
derived using simple 5th-order polynomials, whereas
panels~\ref{figure:pol_spl}c and~\ref{figure:pol_spl}d display the fits
obtained employing adaptive splines with \mbox{$N_{\mbox{\scriptsize
knots}}=5$}. The impact of modifying the asymmetry coefficient $\xi$ is
explored in panels~\ref{figure:pol_spl}a and~\ref{figure:pol_spl}c (in these
fits, \mbox{$\alpha=2$} and \mbox{$N_{\mbox{\scriptsize maxiter}}=1000$} have
been used; the adaptive splines fits were refined \mbox{$N_{\mbox{\scriptsize
refine}}=10$} times). The dashed blue lines indicate the ordinary 
least-squares fits, i.e., those obtained when there is no effective asymmetry
\mbox{($\xi=1$)}, which in each case was used as the initial guess fit in the
numerical minimisation process. For relatively low values of the asymmetry
coefficient (\mbox{$\xi=10$} or~100) the fits are not as good as when using the
largest value \mbox{($\xi=1000$)}.  This is easy to understand, since the
relatively large number of points to be fitted in this example
\mbox{($N=3847$)}, requires that the points that still fall in the outer side
of the boundary during the numerical minimisation of Eq.~(\ref{equation:gls})
overcome the pulling effect of the points in the inner side of the boundary. On
the other hand, panels~\ref{figure:pol_spl}b and~\ref{figure:pol_spl}d display
the effect of changing the power $\alpha$ in the fits. Again, the dashed blue
lines correspond to the ordinary least-squares fits (in the rest of the cases
\mbox{$\xi=1000$} and \mbox{$N_{\mbox{\scriptsize maxiter}}=1000$} have been
used; the adaptive splines fits were refined \mbox{$N_{\mbox{\scriptsize
refine}}=10$} times). In these cases, the best boundary fits are obtained for
\mbox{$\alpha=1$}, whereas for the larger values the fits depart from the
expected result.

\begin{figure}
\includegraphics[angle=-90,width=\columnwidth]{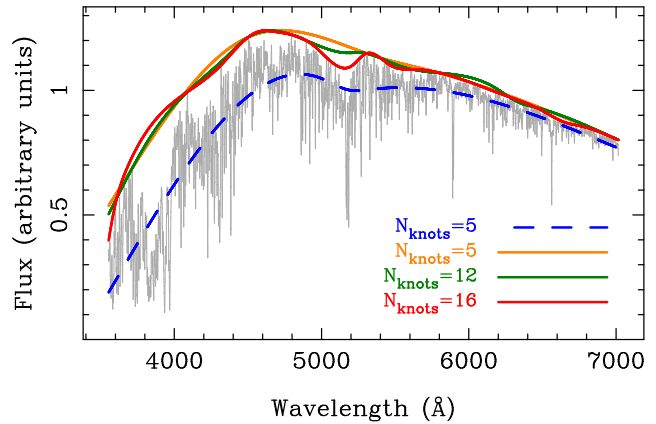}
\caption{Examples of pseudo-continuum fits obtained using adaptive splines with
different number of knots. The same stellar spectrum displayed in
Fig.~\ref{figure:pol_spl} is employed here. The dashed blue line indicates
the ordinary least-squares fit of the data ($\xi=1$, $\alpha=2$). In the rest
of the fits, \mbox{$\xi=1000$}, \mbox{$\alpha=1$} and
\mbox{$N_{\mbox{\scriptsize refine}}=20$} have been used. The effect of using a
different value of \mbox{$N_{\mbox{\scriptsize knots}}$} is clearly visible.
See discussion in Section~\ref{section:application_pseudo}.}
\label{figure:spl_knots}
\end{figure}

The above example illustrates that the optimal asymmetry coefficient $\xi$ and
power $\alpha$ during the boundary procedure can (and must) be tuned for the
particular problem under study. Not surprisingly, this fact also concerns the
number of knots when using adaptive splines.  Fig.~\ref{figure:spl_knots}
shows the different results obtained when estimating the pseudo-continuum in
the same stellar spectrum previously considered, employing different values of
\mbox{$N_{\mbox{\scriptsize knots}}$}. As expected, the fit adapts to the
irregularities exhibited by the spectrum as the number of knots increases. This
is something that for some purposes may not be desired. For instance, the fits
obtained with \mbox{$N_{\mbox{\scriptsize knots}}=12$}, and more notably with
\mbox{$N_{\mbox{\scriptsize knots}}=16$}, detect the absorption around the
Mg~{\sc i} feature at $\lambda\sim5200$~\AA, and for this reason these fits
underestimate the total absorption produced at this wavelength region. In
situations like this the boundary obtained with a lower number of knots may be
more suitable. Obviously there is no general rule to define the right
$N_{\mbox{\scriptsize knots}}$, since the most convenient value will depend on
the nature of the problem under study.

\begin{figure*}
\includegraphics[angle=-90,width=\textwidth]{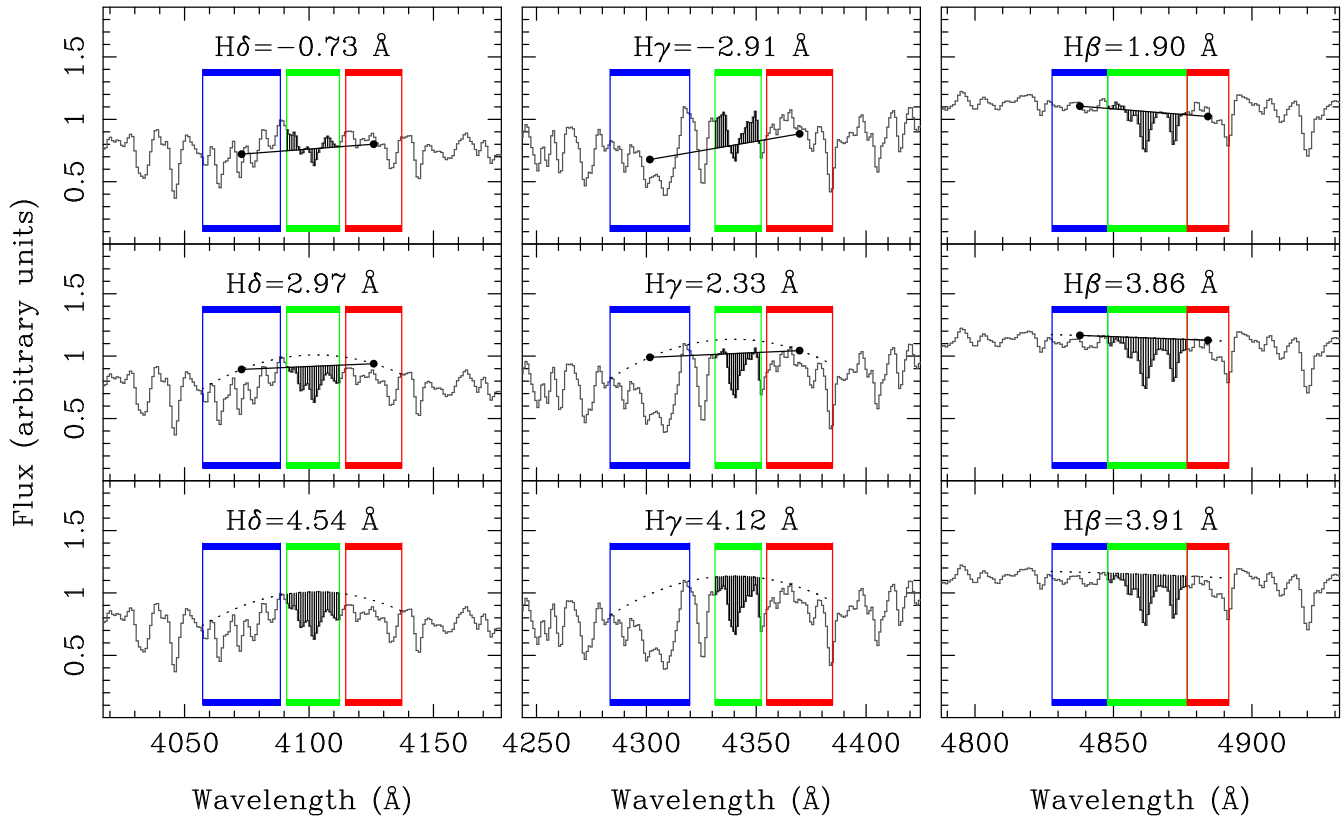}
\caption{Comparison of different strategies in the computation of the
pseudo-continuum for the measurement of line-strength indices. The same stellar
spectrum displayed in Fig.~\ref{figure:pol_spl} is employed here. In this
example three Balmer features are analised, namely H$\delta$, H$\gamma$ and
H$\beta$ (from left to right), showing the commonly employed blue, central and
red sidebands used in their measurement. Top panels correspond to the
traditional method in stellar population studies, in which the pseudo-continuum
is computed as the straight line joining the mean fluxes in the blue and red
sidebands, respectively. In the middle panels the pseudo-continua have been
computed as the straight line joining the values of the upper boundary fits
(second order polynomials fitted to the three bandpasses; dotted lines),
evaluated at the centres of the blue and red bandpasses.  Finally, in the
bottom panels the pseudo-continua are not computed as straight lines, but as
the upper boundary fits themselves.  In each case the resulting line-strength
value (area comprised between the pseudo-continuum fit and the stellar
spectrum) is shown. See discussion in
Section~\ref{section:application_pseudo}.}

\label{figure:balmer_lines}
\end{figure*}

In order to obtain a quantitative determination of the impact of using the
upper boundary fit instead in the estimation of local pseudo-continuum,
Fig.~\ref{figure:balmer_lines} compares the actual line-strength indices
derived for three Balmer lines (H$\beta$, H$\gamma$ and H$\delta$, from right
to left) using three different strategies. For this particular example the same
stellar spectrum displayed in Fig.~\ref{figure:pol_spl} has been used.
Overplotted on each spectrum are the bandpasses typically used for the
measurement of these spectroscopic features. In particular, de bandpasses
limits for H$\beta$ are the revised values given by \citet{trager97}, whereas
for H$\gamma$ and H$\delta$ the limits correspond to H$\gamma_F$ and
H$\delta_F$, as defined by \citet{worthey97}. For each feature, the
corresponding line-strength has been computed by determining the
pseudo-continuum using: i) the straight line joining the mean fluxes in the
blue and red bandpasses (top panels) which is the traditional method; ii) the
straight line joining the values of the upper boundary fits evaluated at the
centres of the same bandpasses (central panels); and iii) the upper boundary
fits themselves (bottom panels). For the cases ii) and iii) the upper boundary
fits have been derived using a second order polynomial fitted to the three
bandpasses. The resulting line-strength indices, numerically displayed above
each spectrum, have been computed as the area comprised between the adopted
pseudo-continuum fit and the stellar spectrum within the central bandpass. For
the three Balmer lines it is clear that the use of the boundary fit provides
larger indices. The traditional method provides very bad values for H$\gamma$
and H$\delta$ (which are even negative!), given that the pseudo-continuum is
very seriously affected by the absorption features in the continuum bandpasses.
This is a well-known problem that has led many authors to seek for alternative
bandpass definitions \citep[see e.g.][]{rose94,vazdekis99} which, on the
other hand, are not immune to other problems related to their sensitivity to
spectral resolution and their high signal-to-noise requirements. These are very
important issues that deserve a much careful analysis, that is beyond the aim
of this paper, and they are going to be studied in a forthcoming work (Cardiel
2009, in preparation).

\begin{figure*}
\includegraphics[angle=-90,width=\textwidth]{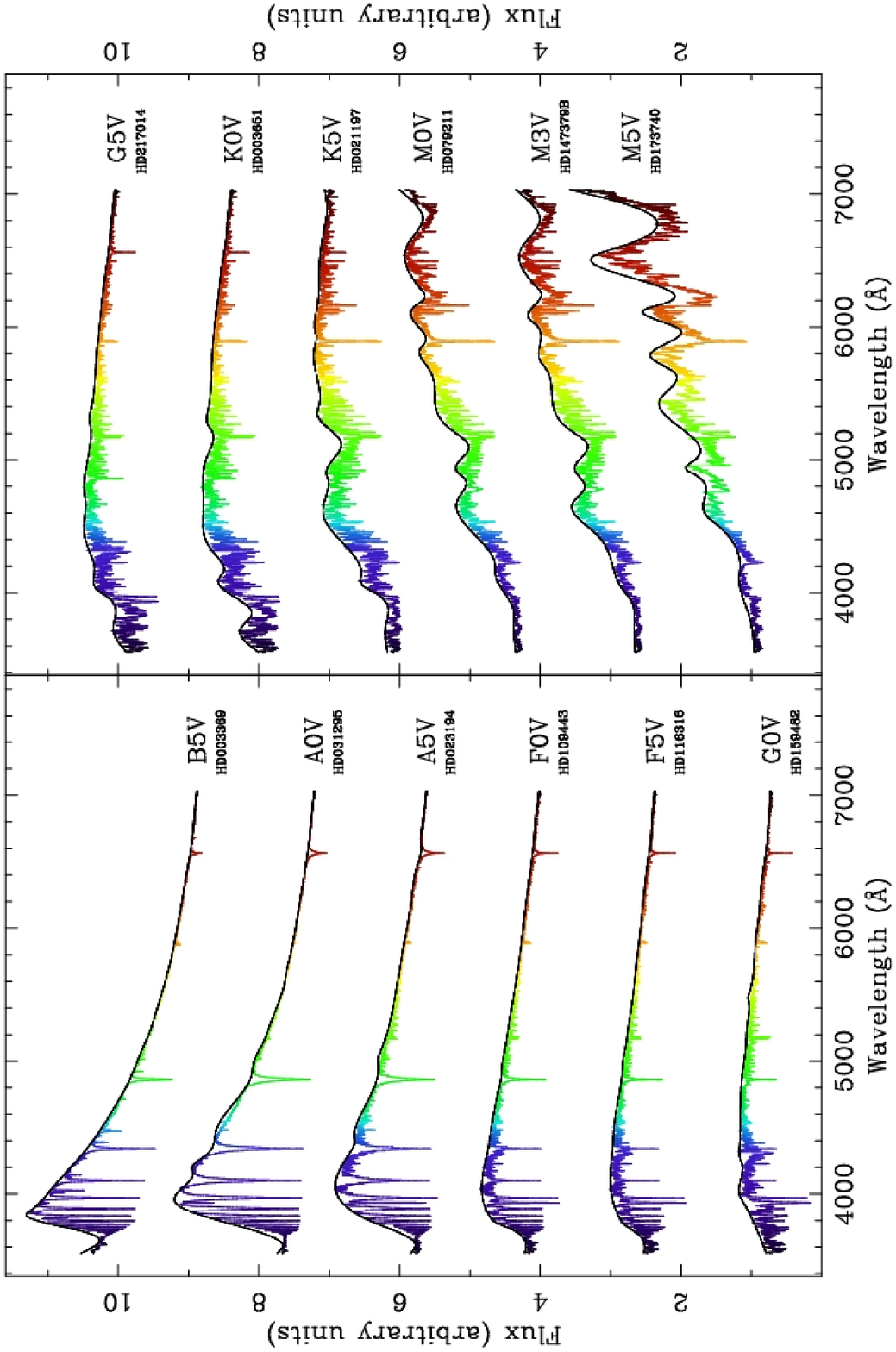}
\caption{Examples of pseudo-continuum fits using adaptive splines. Several
stars from the stellar library MILES \citep{miles}, spanning different spectral
types, have been selected. The fitted pseudo-continua (continuous black line) 
have been automatically determined employing
\mbox{$N_{\mbox{\scriptsize knots}}=19$},
\mbox{$N_{\mbox{\scriptsize maxiter}}=1000$}, \mbox{$N_{\mbox{\scriptsize
refine}}=20$}, \mbox{$\xi=1000$}, \mbox{$\alpha=2$} and \mbox{$\beta=0$}.}
\label{figure:miles}
\end{figure*}

The results of Fig.~\ref{figure:pol_spl} reveal that, for the wavelength
interval considered in that example, the boundary determinations obtained by
using polynomials and adaptive splines are not very different. However, it is
expected that as the wavelength range increases and the expected
pseudo-continuum becomes more complex, the larger flexibility of adaptive
splines in comparison with simple polynomials should provide better fits. To
explore this flexibility in more detail, Fig.~\ref{figure:miles} shows the
result of using adaptive splines to estimate the pseudo-continuum of 12
different spectra corresponding to stars exhibiting a wide range of spectral
types (from B5V to M5V), selected from the empirical stellar library MILES
\citep{miles} previously mentioned. Although in all the cases the fits have
been computed blindly without considering the use of an initial knot
arrangement appropriate for the particularities of each spectral type, it is
clear from the figure that adaptive splines are flexible enough to give
reasonable fits independently of the considered star. More refined fits can be
obtained using an initial knot pattern more adjusted to the curvature of the
pseudo-continuum exhibit by the stellar spectra.

\begin{figure}
\includegraphics[angle=0,width=\columnwidth]{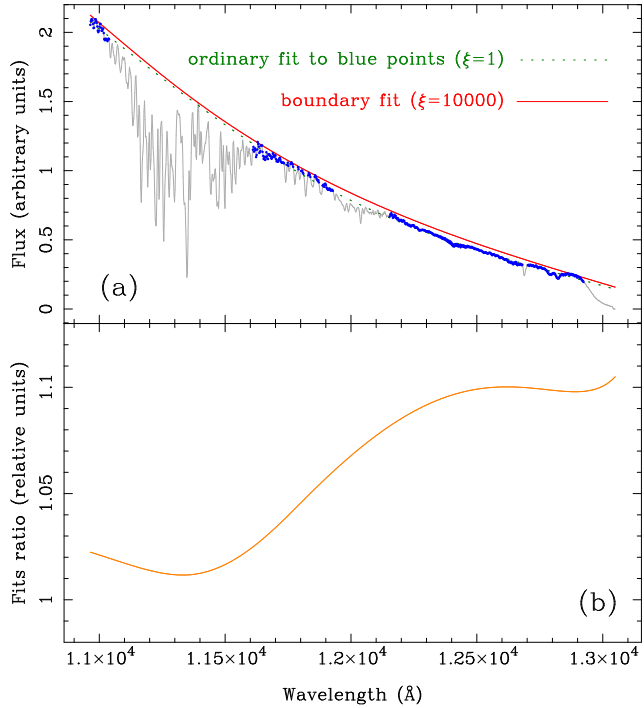}
\caption{Comparison of the results of using an ordinary fit and adaptive
splines when deriving the telluric correction in a particular spectroscopic
calibration.  \emph{Panel~(a)}: the light grey line corresponds to the spectrum
obtained in the \emph{J}~band of the hot star HD165174. Some special points of
this spectrum have been manually selected (small blue points) to determine the
approximate pseudo-continuum. The resulting ordinary fit to adaptive splines
(i.e.\ adopting $\xi=1$) using exclusively these selected points is displayed
with the dotted green line. A more suitable fit (continuous red line) is
obtained employing $\xi=10000$, in which case the fit is performed over the
whole spectrum.  The two fits have been carried out with
\mbox{$N_{\mbox{\scriptsize knots}}=3$}, \mbox{$N_{\mbox{\scriptsize
maxiter}}=1000$}, \mbox{$N_{\mbox{\scriptsize refine}}=10$}, \mbox{$\alpha=2$}
and \mbox{$\beta=0$}. \emph{Panel~(b)}: ratio between the two fits displayed in
the previous panel.}
\label{figure:telluric}
\end{figure}

A good estimation of spectra pseudo-continuum is very useful, for example, when
correcting spectroscopic data from telluric absorptions using featureless (or
almost featureless) calibration spectra. This is a common strategy when
performing observations in the near-infrared windows.
Fig.~\ref{figure:telluric}a illustrates a typical example, in which the
observation of the hot star V986~Oph (HD165174, spectral type B0III) is
employed to determine the correction. This star was observed in the
\emph{J}~band as part of the calibration work of the observations presented in
\citet{cardiel03}. The stellar spectrum is shown in light grey, whereas the
blue points indicate a manual selection of spectrum regions employed to
estimate the overall pseudo-continuum. The dotted green line corresponds to the
ordinary least-squares fit of these points, whereas the red continuous line is
the upper boundary obtained with adaptive splines using
\mbox{$N_{\mbox{\scriptsize knots}}=3$} with an asymmetry coefficient
\mbox{$\xi=10000$}. In Fig.~\ref{figure:telluric}b the ratio between both
fits in represented, showing that there are differences up to a few percent
between these fits. Two kind of errors are present here. In overall the
ordinary least-squares fit underestimates the pseudo-continuum level, which
introduces a systematic bias on the resulting depth of the telluric features
(the whole curve displayed in Fig.~\ref{figure:telluric}b is above~1.0). In
addition, since the selected blue points do include real (although small)
spectroscopic features, there are variations as a function of wavelength of the
above discrepancy. These differences can be important when trying to perform a
high-quality spectrophotometric calibration. It is important to highlight that
an important additional advantage of the boundary fitting is that this method
does not require the masking of any region of the problem spectrum, which
avoids the effort (and the subjectivity) of selecting special points to guide
the fit.

Another important aspect concerning the use of boundary fits for the
determination of the pseudo-continuum of spectra is that this method can
provide an alternative approach for the estimation of the pseudo-continuum flux
when measuring line-strength indices. Instead of using the average fluxes in
bandpasses located nearby the (typically central) bandpass covering the
relevant spectroscopic feature, the mean flux on the upper boundary can be
employed. In this case it is important to take into account that flux
uncertainties will bias the fits towards higher values. Under these situations
the approach described later in Section~\ref{section:uncertainties} can be
employed. Concerning this problem is worth mentioning here the method
presented by \citet{rogers08}, who employ a boosted median continuum to derive
equivalent widths more robustly than using the classic side-band procedure.

%------------------------------------------------------------------------------
\subsection{Estimation of data ranges}
\label{section:application_levels}

\begin{figure}
\includegraphics[angle=0,width=\columnwidth]{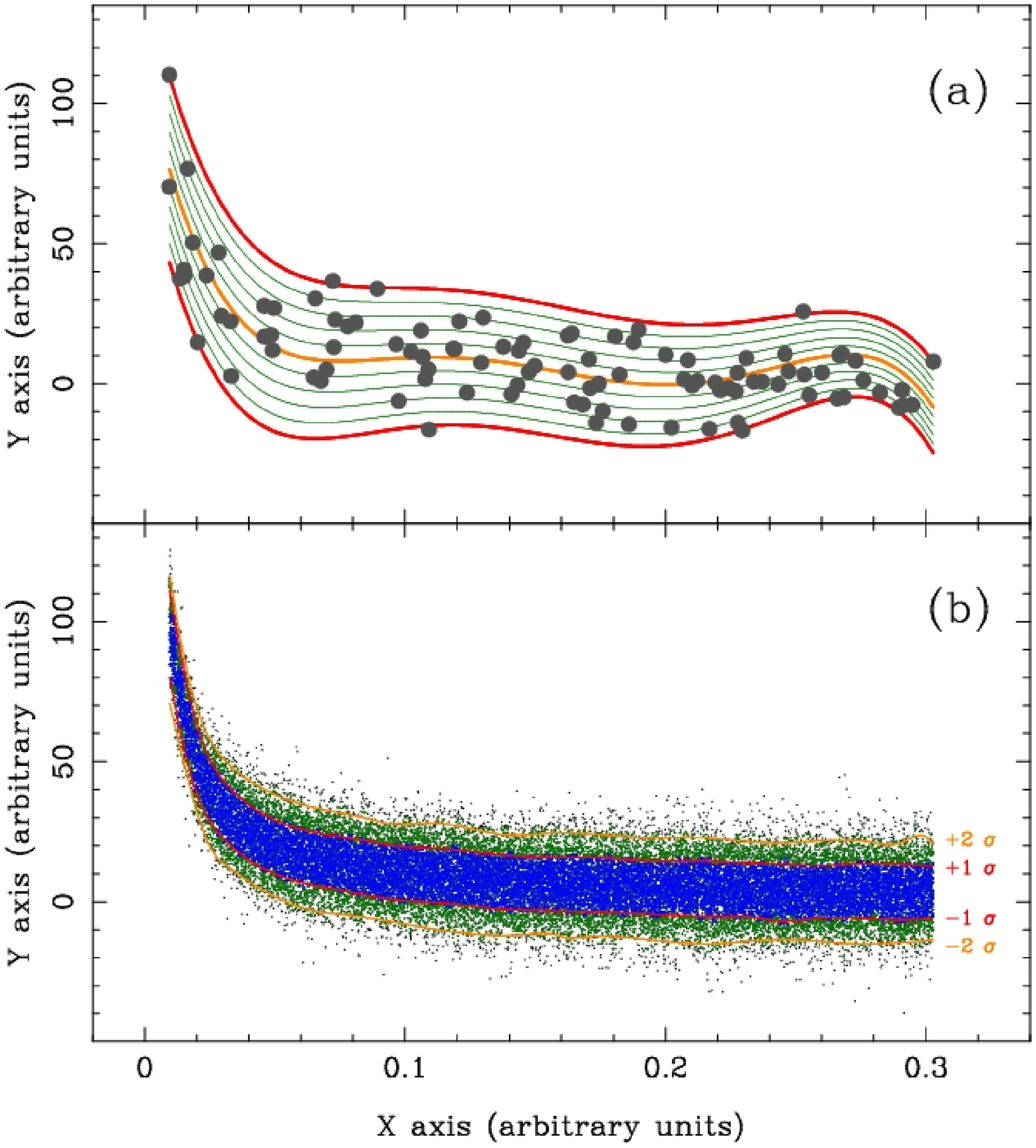}
\caption{Examples of data boundary applications for the estimation of data
ranges. \emph{Panel~(a)}: Using the lower and upper boundary limits for the
data displayed in Figs.~\ref{figure:oneoverx_pol}, \ref{figure:oneoverx_spl}
and~\ref{figure:example}, and computed using simple 5th order polynomials, it
is trivial to subdivide the range spanned by the data in the \mbox{$y$-axis} by
creating a regular grid (i.e.\ contant $\Delta y$ at a fixed $x$) between both
boundary limits. In this example the region has been subdivided in ten
intervals. \emph{Panel~(b)}: 30000 points randomly drawn from the functional
form $y=1/x$, with $\sigma=10$ for all the points.  Splitting the $x$-range in
100~intervals, sorting the data within each interval and keeping track of the
subsets containing 68.27\% ($\pm 1 \sigma$; blue points) and 95.44\% ($\pm 2
\sigma$; green points) of the data points around the median, it is possible to
compute the upper and lower boundaries for those two subsets (continuous red
and orange lines, respectively). The boundaries in this example have been
determined using adaptive splines with \mbox{$N_{\mbox{\scriptsize
knots}}=15$}, \mbox{$N_{\mbox{\scriptsize itermax}}=1000$},
\mbox{$N_{\mbox{\scriptsize refine}}=10$}, \mbox{$\alpha=2$}, and
\mbox{$\beta=0$}.}
\label{figure:levels}
\end{figure}

A quite trivial but useful application of the boundary fits is the empirical
determination of data ranges. One can consider scenarios in which it is needed
to subdivide the region spanned by the data in a particular grid.
Fig.~\ref{figure:levels}a illustrates this situation, making use of the
5th order polynomial boundaries corresponding to the data previously used in
Figs.~\ref{figure:oneoverx_pol}, \ref{figure:oneoverx_spl},
and~\ref{figure:example}. Once the lower and the upper boundaries are
available, it is trivial to generate a grid of lines dividing the region
comprised between the boundaries as needed.

A more complex scenario is that in which the data exhibit a clear scatter
around some tendency, and one needs to determine regions including a
given fraction of the points. A frequent case appears when
one needs to remove outliers, and then it is necessary to obtain an estimation
of the regions containing some relevant percentages of the data. In
Fig.~\ref{figure:levels}b this situation is exemplified with the use of a
simulated data set consisting in 30000 points, for which the regions that
include 68.27\% and 95.44\% of the centred data points, corresponding to $\pm
1\sigma$ and $\pm 2\sigma$ in a normal distribution, have been determined by
first selecting those data subsets, and then fitting their corresponding
boundaries using adaptive splines, as explained with more detail in the figure
caption.

%%%%%%%%%%%%%%%%%%%%%%%%%%%%%%%%%%%%%%%%%%%%%%%%%%%%%%%%%%%%%%%%%%%%%%%%%%%%%%%
\section{The impact of data uncertainties}
\label{section:uncertainties}

Although the method described in Section~\ref{section:the_method} already takes
into account data uncertainties through their inclusion as a weighting
parameter (governed by the exponent $\beta$), it is important to highlight that
this weighting scheme does not prevent the boundary fits to be highly
biased due to the presence of such uncertainties. For example, in the
determination of the pseudo-continuum of a given spectrum, even considering the
same error bars for the fluxes at all wavelengths, the presence of noise
unavoidably produces some scatter around the real data. When fitting the upper
boundary to a noisy spectrum the fit will be dominated by the points that
randomly exhibit the largest positive departures. Under these circumstances,
two different alternatives can be devised:

\begin{figure}
\includegraphics[angle=0,width=\columnwidth]{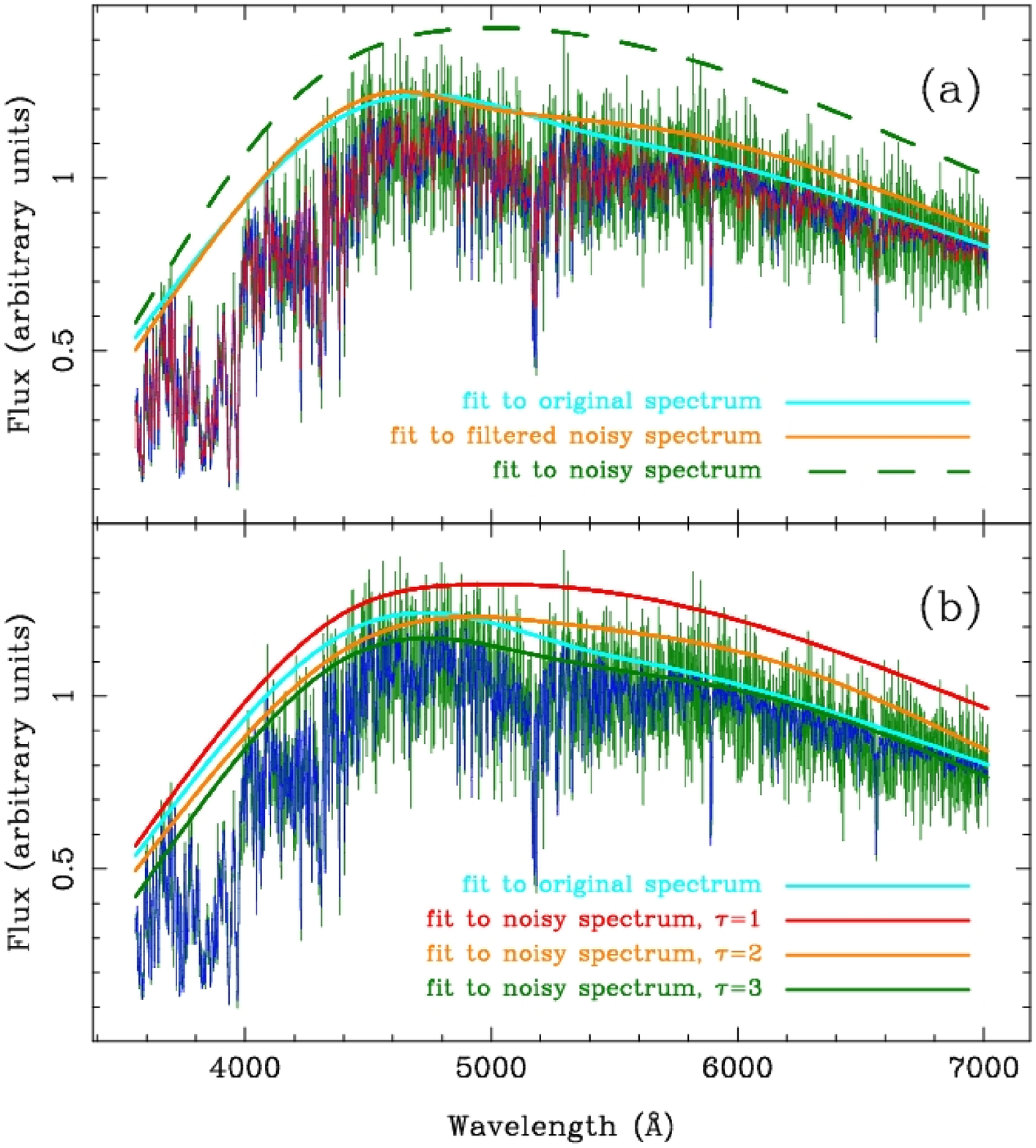}
\caption{Comparison of the two approaches described in
Section~\ref{section:uncertainties} for the boundary fitting with data
uncertainties. \emph{Panel~(a)}: original spectrum of HD003651 without noise
(blue spectrum), spectrum with artificially added noise (green spectrum) and
noisy spectrum after a Gaussian filtering (red spectrum). Note that the
original (blue) and the filtered noisy (red) spectra are almost coincident. The
upper boundary displayed with a dashed green line is the fit to the noisy
spectrum using adaptive splines, whereas the upper boundaries plotted with
continuous orange and cyan lines are the fits to the filtered noisy spectrum
and to the original spectrum, respectively. \emph{Panel~(b)}: original and
noisy spectra are plotted with blue and green lines, respectively (the
filtered spectrum is not plotted here). The cyan line is again the fit to
the original spectrum. The rest of the boundary lines indicate the fits to the
noisy spectrum using different values of the cut-off parameter (red
\mbox{$\tau=1$}, orange \mbox{$\tau=2$}, and green \mbox{$\tau=3$}).  In all
the fits \mbox{$N_{\mbox{\scriptsize knots}}=5$}, \mbox{$N_{\mbox{\scriptsize
maxiter}}=1000$}, \mbox{$\xi=1000$}, \mbox{$\alpha=1$}, \mbox{$\beta=0$}, and
\mbox{$N_{\mbox{\scriptsize refine}}=10$} have been employed. See discussion in
Section~\ref{section:uncertainties}.}
\label{figure:spl_errors1}
\end{figure}

\begin{enumerate}
  \item \emph{To perform a previous rebinning or filtering of the data} prior
  to the boundary fitting, in order to eliminate, or at least minimize, the
  impact of data uncertainties. After the filtering one assumes that these
  uncertainties are not seriously biasing the boundary fit. In this way one can
  employ the same technique described in Section~\ref{section:the_method}. This
  approach is illustrated in Fig.~\ref{figure:spl_errors1}a. In this case the
  original spectrum of HD00365 (also employed in Figs.~\ref{figure:pol_spl}
  and~\ref{figure:spl_knots}), as extracted from the MILES library
  \citep{miles}, is considered as a noise-free spectrum (plotted in blue). Its
  corresponding upper boundary fit using adaptive splines with
  \mbox{$N_{\mbox{\scriptsize knots}}=5$} is shown as the cyan line. This
  original spectrum has been artificially degraded by considering an arbitrary
  signal-to-noise ratio per pixel \mbox{S/N=10} (displayed in green), and the
  resulting upper boundary fit is shown with a dashed green line. It is obvious
  that this last fit is highly biased, being dominated by the points with
  higher fluxes. Finally, the noisy spectrum has been filtered by convolving it
  with a Gaussian kernel (of standard deviation 100~km/s), with the result
  being over-plotted in red. Note that this filtered spectrum overlaps almost
  exactly with the original spectrum. The boundary fit plotted with the
  continuous orange line is the upper boundary for that filtered spectrum.
  Although the result is not the same as the one derived with the original
  spectrum, it is much better than the one directly obtained over the noisy
  spectrum.

  \item \emph{To allow a loose boundary fitting.} Another possibility consists
  in trying to leave a fraction of the points with extreme values to fall
  outside (i.e., in the wrong side) of the boundary, specially those with
  higher uncertainties. This option is easy to parametrize by introducing a
  cut-off parameter $\tau$ into the overall weighting factors given in
  Eq.~(\ref{equation:asymmetry}). The new factors can then be computed as
  \begin{equation}
  w_i \equiv \left\{
  \begin{array}{ll}
  \begin{array}{@{}c@{}}\textrm{upper}\\ \textrm{boundary}\end{array} &
      \left\{ \begin{array}{ll}
                    1/\sigma_i^\beta & \textrm{for}\;\; y(x_i) \ge y_i
                      -\tau \sigma_i\\
                  \xi/\sigma_i^\beta & \textrm{for}\;\; y(x_i) < y_i
                      -\tau \sigma_i
       \end{array} \right. \\
                          & \\
  \begin{array}{@{}c@{}}\textrm{lower}\\ \textrm{boundary}\end{array} &
       \left\{ \begin{array}{ll}
                  \xi/\sigma_i^\beta & \textrm{for}\;\; y(x_i) > y_i
                      +\tau \sigma_i\\
                    1/\sigma_i^\beta & \textrm{for}\;\; y(x_i) \le y_i
                      +\tau \sigma_i
       \end{array} \right.
  \end{array}
  \right.
  \label{equation:asymmetry_errors}
  \end{equation}
  where $\sigma_i$ is the uncertainty associated to the dependent variable
  $y_i$. The cut-off parameter assigns to a point that falls outside of the
  boundary by distance that is less than or equal to $\tau\sigma_i$ the same
  low weight during the fitting procedure than the weight that receive the
  inner points.  In other words, points like that do not receive the extra
  weighting factor provided by the asymmetry coefficient $\xi$, even though
  they are outside of the boundary. Note that \mbox{$\tau=0$} simplifies the
  algorithm to the one described in Section~\ref{section:the_method}.
  Fig.~\ref{figure:spl_errors1}b illustrates the use of the cut-off parameter
  $\tau$ in the upper boundary fitting of the spectrum of HD003651. The
  cyan boundary is again the upper boundary determination using adaptive
  splines with the original spectrum. The rest of the boundary fits correspond
  to the use of the weighting scheme given in
  Eq.~(\ref{equation:asymmetry_errors}) for different values of $\tau$, as
  indicated in the legend. As $\tau$ increases, a larger number of points are
  left outside of the boundary during the minimisation procedure. In the
  example, the value \mbox{$\tau=3$} seems to give a reasonable fit in the
  redder part of the spectrum, although in the bluer region the corresponding
  fit is too low. It is clear from this example that to define a \emph{correct}
  value of $\tau$ is not a trivial issue. Most of the times the most suited
  $\tau$ will be a compromise between a high value (in order to avoid the bias
  introduced by highly deviant points) and a low value (in order to avoid
  leaving outside of the boundary right data points).
\end{enumerate}

\begin{figure}
\includegraphics[angle=0,width=\columnwidth]{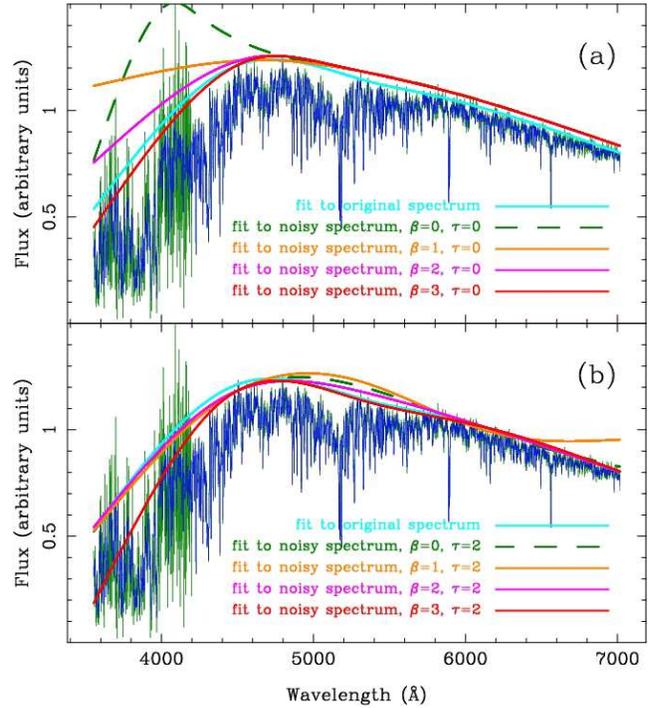}
\caption{Study of the impact of variable signal-to-noise ratio in the upper
boundary fitting of the spectrum of the star HD003651. In both panels the
original spectrum (blue) is plotted together with the same spectrum after
artificially adding noise (green) corresponding to a signal-to-noise ratio per
pixel S/N=3 for \mbox{$\lambda \leq 4200$~\AA}, and to S/N=50 for
\mbox{$\lambda > 4200$~\AA}. The cyan line indicates the upper boundary fit to
the original spectrum. \emph{Panel~(a)}: In these fits the cut-off parameter
has been ignored (\mbox{$\tau=0$}), but different values of the power $\beta$,
as indicated in the legend, are employed. Note that the unweighted fit
(\mbox{$\beta=0$}; dashed green line) is highly biased. \emph{Panel~(b)}: the
same fits of the previous panel are repeated here but using \mbox{$\tau=2$}. In
all the fits \mbox{$N_{\mbox{\scriptsize knots}}=5$},
\mbox{$N_{\mbox{\scriptsize maxiter}}=1000$}, \mbox{$\xi=1000$},
\mbox{$\alpha=1$}, and \mbox{$N_{\mbox{\scriptsize refine}}=10$} have been
employed. See discussion in Section~\ref{section:uncertainties}.}
\label{figure:spl_errors2}
\end{figure}

An additional complication arises when one combines in the same data set points
with different uncertainties. It is in these situations when the role of the
power $\beta$ in Eq.~(\ref{equation:gls}) becomes important. To illustrate the
situation, Fig.~\ref{figure:spl_errors2} shows the different pseudo-continuum
estimations obtained again for the star HD003651, but now considering that the
spectrum is much noisier below 4200~\AA\ than above this wavelength. In
panel~\ref{figure:spl_errors2}a the fits are derived ignoring the cut-off
parameter previously discussed (i.e.\ assuming \mbox{$\tau=0$}), but with
different values of $\beta$. In the unweighted case (\mbox{$\beta=0$}, dashed
green line) the resulting upper boundary is dramatically biased for
$\lambda<4200$~\AA\ due to the presence of highly deviant fluxes. The use of
non-null (and positive) values of $\beta$ induces the fit to be less dependent
on the noisier values, being necessary a value as high as $\beta=3$ to obtain a
fit similar to the one obtained in absence of noise (cyan line). However, since
the fitted spectrum (green) do still have noise for $\lambda>4200$~\AA, all the
fits in that region are still biased compared to the fit for the original
spectrum (cyan). In order to deal not only with the variable noise, but with
the noise itself independently of its absolute value, it is possible to combine
the effect of a tuned $\beta$ value with the introduction of a cut-off
parameter $\tau$.  Fig.~\ref{figure:spl_errors2}b shows the results derived
employing a fixed value \mbox{$\tau=2$} with the same variable values of
$\beta$ used in the previous panel. In this case, the boundary corresponding to
$\beta=2$ (magenta) exhibits an excellent agreement with the fit for the
original spectrum (cyan) at all wavelengths. Thus, the combined effect of an
error-weighted fit and the use of a cut-off parameter is providing a reasonable
boundary determination, even under the presence of wavelength dependent noise.

%%%%%%%%%%%%%%%%%%%%%%%%%%%%%%%%%%%%%%%%%%%%%%%%%%%%%%%%%%%%%%%%%%%%%%%%%%%%%%%
\section{Conclusions}
\label{section:conclusions}

This work has confronted the problem of obtaining analytical expressions for
the upper and lower boundaries of a given data set. The task reveals treatable
using a generalised version of the very well-known ordinary least-squares fit
method. The key ideas behind the proposed method can be summarised as follows:
\begin{itemize}
  \item The sought boundary is iteratively determined starting from an initial
  guess fit. For the analysed cases an ordinary least-squares fit provides a
  suitable starting point. At every iteration in the procedure a particular fit
  is always available.
  \item In each iteration the data to be fitted are segregated in two subgroups
  depending on their position relative to the particular fit at that iteration.
  In this sense, points are classified as being inside or outside of the
  boundary.
  \item Points located outside of the boundary are given an extra weight in the
  cost function to be minimized. This weight is parametrized through the
  \emph{asymmetry coefficient} $\xi$. The net effect of this coefficient is to
  generate a stronger pulling effect of the outer points over the fit, which in
  this way shifts towards the frontier delineated by the outer points as the
  iterations proceed.
  \item The distance from the points to a given fit are introduced in the cost
  function with a variable power $\alpha$, not necessarily in the traditional
  squared way. This supplies an additional parameter to play with when
  performing the boundary determination.
  \item Since data uncertainties are responsible for the existence of highly
  deviant points in the considered data sets, their incorporation in the
  boundary determination has been considered in two different and complementary
  ways. Errors can readily be incorporated into the cost function as weighting
  factors with a variable power $\beta$ (which does not have to be necessarily
  two). In addition, a cutt-off parameter $\tau$ can also be tuned to exclude
  outer points from receiving the extra factor given by the asymmetry
  coefficient depending on the absolute value of their error bar.  The use of
  both parameters ($\beta$ and $\tau$) provides enough flexibility to handle
  the role of the data uncertainties in different ways depending on the nature
  of the considered boundary problem. 
  \item The minimisation of the cost function can be easily carried out using
  the popular {\sc downhill} simplex method. This allows the use of any
  computable function as the analytical expression for the boundary fits.
\end{itemize}

The described fitting method has been illustrated with the use of simple
polynomials, which probably are enough for most common situations. For those
scenarios where the data exhibit rapidly changing values, a more powerful
approach, using \emph{adaptive splines}, has also been described.  Examples
using both simple polynomials and adaptive splines have been presented, showing
that they are good alternatives to estimate the pseudo-continuum of spectra and
to segregate data in ranges.

The analysed examples have shown that there is no magic rule to \emph{a priori}
establish the most suitable values for the tunable parameters ($\xi$, $\alpha$,
$\beta$, $\tau$, $N_{\mbox{\scriptsize maxiter}}$, $N_{\mbox{\scriptsize
knots}}$). The most appropriate choices must be accordingly tuned for the
particular problem under study. In any case, typical values for some of these
parameters in the considered examples are $\xi\in[1000,10000]$ and
$\alpha\in[1,2]$. Unweighted fits require $\beta=0$. To take into account data
uncertainties one must play around with the $\beta$ and $\tau$ parameters
(which typical values range from~0 to~3).

A new program called {\tt BoundFit} (and available at the URL given in
Section~\ref{section:introduction}) has been written by the author to help any
person interested in playing with the method described in this paper. It is
important to note that for some problems it is advisable to normalise the data
ranges prior to the fitting computation in order to prevent (or at least
reduce) numerical errors.  {\tt BoundFit} incorporates this option, and the
users should verify the benefit of applying such normalisation for their
particular needs.

%%%%%%%%%%%%%%%%%%%%%%%%%%%%%%%%%%%%%%%%%%%%%%%%%%%%%%%%%%%%%%%%%%%%%%%%%%%%%%%
\section*{Acknowledgements}

Valuable discussions with Guillermo Barro, Juan Carlos Mu\~{n}oz and
Javier Cenarro are gratefully acknowledged. The author is also grateful to the
referee, Charles Jenkins, for his useful comments. This work was supported by
the Spanish Programa Nacional de Astronom\'{\i}a y Astrof\'{\i}sica under grant
\mbox{AYA2006--15698--C02--02}.

%%%%%%%%%%%%%%%%%%%%%%%%%%%%%%%%%%%%%%%%%%%%%%%%%%%%%%%%%%%%%%%%%%%%%%%%%%%%%%%
%%%%%%%%%%%%%%%%%%%%%%%%%%%%%%%%%%%%%%%%%%%%%%%%%%%%%%%%%%%%%%%%%%%%%%%%%%%%%%%
\appendix

\section{Introducing additional constraints in the fits}
\label{appendix:constraints}

Sometimes it is not only necessary to obtain a given functional fit to a data
set, but to do so while imposing restrictions on some of the fitted parameters
\mbox{$a_0,a_1,\ldots,a_p$}. This can be done by introducing either equality or
inequality constraints, or both. These constraints are normally expressed as
\begin{eqnarray}
c_j(a_0,a_1,\ldots,a_p) = 0  & &
j=1,\ldots,n_{\mbox{\scriptsize e}}
\label{equation:equality_constraint} \\
c_j(a_0,a_1,\ldots,a_p)\geq 0 & &
j=n_{\mbox{\scriptsize e}}+1,\ldots,
n_{\mbox{\scriptsize e}}+n_{\mbox{\scriptsize i}}
\label{equation:inequality_constraint}
\end{eqnarray}
being $n_{\mbox{\scriptsize e}}$ and $n_{\mbox{\scriptsize i}}$ the number of
equality and inequality constraints, respectively. In the case of some boundary
determinations it may be useful to incorporate these type of
constraints, for example when one needs the boundary fit to pass through some
pre-defined fixed points, and/or to have definite derivatives at some
points (allowing for a smooth connection between functions).

Many techniques that allow to minimize cost functions while taking into account
supplementary constraints are described in the literature \citep[see
e.g.][]{optimization_rao, optimization_gill, optimization_bazaraa,
optimization_nocedal, optimization_fletcher}, and to explore them here in
detail are beyond the aim of this work. However this appendix outlines two
basic approaches that can be useful for some particular situations.

%------------------------------------------------------------------------------
\subsection{Avoiding the constraints}

Before facing the minimisation of a constrained fit, it is advisable to check
whether some simple transformations can help to convert the constrained
optimisation problem into an unconstrained one by making change of variables.
\citet{optimization_rao} presents some useful examples. For instance, a
frequently encountered constraint is that in which a given parameter $a_l$ is
restricted to lie within a given range, e.g.  \mbox{$a_{l,\mbox{\scriptsize
min}} \leq a_l \leq a_{l,\mbox{\scriptsize max}}$}. In this case the simple
transformation
\begin{equation}
a_l = a_{l,\mbox{\scriptsize min}} +
(a_{l,\mbox{\scriptsize max}}-a_{l,\mbox{\scriptsize min}}) \sin^2 b_l
\end{equation}
provides a new variable $b_l$ which can take any value. If the original
parameter is restricted to satisfy $a_l > 0$, the trivial transformations
$a_l=\mbox{abs}(b_l)$, $a_l=b_l^2$, or $a_l=\exp(b_l)$ can be useful.

Unfortunately, when the constraints are not simple functions, it is not
easy to find the required transformations. As highlighted by
\citet{optimization_fletcher}, the transformation procedure is not always free
of risk, and in the case where it is not possible to eliminate all the
constraints by making change of variable, it is better to avoid partial
transformation \citep{optimization_rao}.

An additional strategy that can be employed when handling equality constraints
is trying to use the equations to eliminate some of the variables. For example,
if for a given equality constraint $c_j$ is possible to rearrange the
expression to solve for one of the variables
\begin{equation}
c_j=0 \longrightarrow a_s = g_j(a_0,a_1,\ldots,a_{s-1},a_{s+1},\ldots,a_p),
\end{equation}
then the cost function simplifies from a function in $(p+1)$ variables into a
function in $p$ variables
\begin{equation}
\begin{array}{@{}l}
f(a_0,a_1,\ldots,a_{s-1},a_s,a_{s+1}\ldots,a_p) = \\
=f(a_0,a_1,\ldots,a_{s-1},g_j,a_{s+1}\ldots,a_p),
\end{array}
\end{equation}
since the dependence on $a_s$ is removed. When the considered problem only has
equality constraints and, in addition, for all of them it is possible to apply
the above elimination, the fitting procedure transforms into a simpler
unconstrained problem.
 
%------------------------------------------------------------------------------
\subsection{Facing the constraints}

The weighting scheme underlying the minimisation of Eq.~(\ref{equation:gls}) is
actually an optimisation process based on the penalisation in the cost function
of the data points that falls in the \emph{wrong} side (i.e.\ outside) of the
boundary to be fitted. For this reason it seems appropriate to employ
additional penalty functions \citep[see e.g.][]{optimization_bazaraa} to
incorporate constraints into the fits.

In the case of constraining the range of some of the parameters to be fitted,
\mbox{$a_{l,\mbox{\scriptsize
min}} \leq a_l \leq a_{l,\mbox{\scriptsize max}}$},
it is trivial to adjust the value of the cost function by introducing a large
factor $\Lambda$ that clearly penalises parameters beyond the required limits.
In this sense, Eq.~(\ref{equation:gls}) can be rewritten as
\begin{equation}
f=
\Lambda h(a_0,a_1,\ldots,a_p) +
\sum_{i=1}^{N} w_i | y(x_i)-y_i |^\alpha.
\label{equation:gls_constrained}
\end{equation}
where $h(a_0,a_1,\ldots,a_p)$ is a function that is null when the required
parameters are within the requested ranges (i.e., the fit is performed in an
unconstrained way), and some positive large value for the contrary situation. 

For the particular case of equality constraints of the form given in
Eq.~(\ref{equation:equality_constraint}), it is possible to directly
incorporate these constraints into the cost functions as
\begin{equation}
f=
\Lambda \sum_{j=1}^{n_{\mbox{\scriptsize e}}}|c_j(a_0,a_1,\ldots,a_p)|^\alpha +
\sum_{i=1}^{N} w_i | y(x_i)-y_i |^\alpha.
\label{equation:penalized_cost}
\end{equation}
In this situation, for the constraints to have an impact in the cost function,
the value of the penalisation factor $\Lambda$ must be large enough to
guarantee that the first summation in Eq.~(\ref{equation:penalized_cost})
dominates over the second summation when a temporary solution implies a large
value for any $|c_j|$.

\begin{figure}
\includegraphics[angle=0,width=\columnwidth]{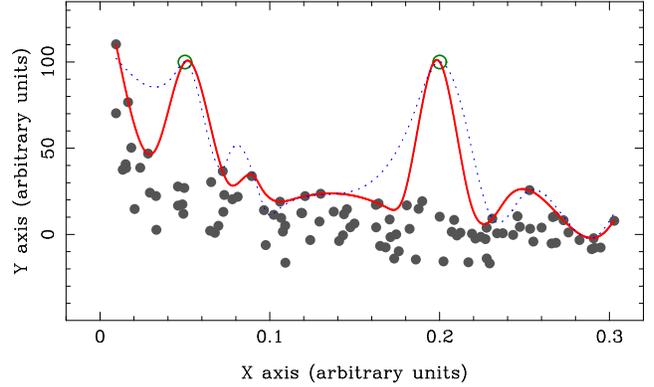}
\caption{Example of constrained boundary fit, using adaptive splines with the
same data employed in Figs.~\ref{figure:oneoverx_pol},
\ref{figure:oneoverx_spl} and~\ref{figure:example}. The boundary (red line) has
been forced to pass through the points marked with open circles (green), namely
(0.05,100) and (0.20,100). To give an important weight to the two constraints
in Eq.~(\ref{equation:penalized_cost}), the value of the penalisation factor
has been set to \mbox{$\Lambda=10^6$}. The dotted blue line is the same fit,
but introducing two new additional constraints, in particular forcing the
derivatives to be zero at the same fixed points.}
\label{figure:constraints}
\end{figure}

As an example, Fig.~\ref{figure:constraints} displays the upper boundary
limit computed using adaptive splines for the same data previously employed in
Figs.~\ref{figure:oneoverx_pol}, \ref{figure:oneoverx_spl}
and~\ref{figure:example}, but arbitrarily forcing the fit to pass through the
two fixed points (0.05,100) and (0.20,100), marked in the figure with the green
open circles. The constrained fit (thick continuos red line) has been
determined by introducing the two equality constraints
\begin{equation}
\begin{array}{@{}ll}
c_1: & y(x=0.05)-100=0, \quad \mbox{and} \\
c_2: & y(x=0.20)-100=0. 
\end{array}
\end{equation}
The displayed fit was computed using a penalisation factor $\Lambda=10^6$,
with an asymmetry coefficient \mbox{$\xi=1000$}, \mbox{$N_{\mbox{\scriptsize
knots}}=15$}, \mbox{$N_{\mbox{\scriptsize maxiter}}=1000$} iterations,
\mbox{$N_{\mbox{\scriptsize refine}}=20$} processes, \mbox{$\alpha=2$}, and
\mbox{$\beta=0$}. For comparison, another fit (dotted blue line) has also
been computed by introducing two more constraints, namely forcing the
derivatives to be zero at the same points, i.e., \mbox{$y'(x=0.05)=0$} and
\mbox{$y'(x=0.20)=0$}. The resulting fit is clearly different, highlighting the
importance of the introduction of the constraints.

%%%%%%%%%%%%%%%%%%%%%%%%%%%%%%%%%%%%%%%%%%%%%%%%%%%%%%%%%%%%%%%%%%%%%%%%%%%%%%%

\section[]{Normalisation of data ranges to reduce numerical errors}
\label{appendix:normalization}

The appearance of numerical errors is one of the most important sources of
problems when fitting functions, in particular polynomials, to any data set
making use of a piece of software. The problems can be specially serious when
handling large data sets, using high polynomial degrees, and employing
different and large data ranges. Since the size of the data set is usually
something that one does not want to modify, and the polynomial degree is also
fixed by the nature of the data being modelled (furthermore in the case of cubic
splines, where the polynomial degree is fixed), the easier way to reduce the
impact of numerical errors is to normalise the data ranges prior to the
fitting procedure. However, although this normalisation is a straightforward
operation, the fitted coefficients cannot be directly employed to evaluate the
sought function in the original data ranges. Previously it is necessary to
properly transform those coefficients. This appendix provides the corresponding
coefficient transformations for the case of the fitting to simple
one-dimensional polynomials and to cubic splines.

%------------------------------------------------------------------------------
\subsection{Simple polynomials}

Simple polynomials are typically expressed as
\begin{equation}
y = a_0+a_1 x + a_2 x^2 + \cdots + a_p x^p.
\label{equation_fitted_polynomial}
\end{equation}
Let's consider that the ranges exhibited by the data in the corresponding
coordinate axes are given by the intervals
$[x_{\mbox{\scriptsize min}},x_{\mbox{\scriptsize max}}]$ and
$[y_{\mbox{\scriptsize min}},y_{\mbox{\scriptsize max}}]$, and assume that
one wants to normalise the data within these intervals into new ones given by
$[\tilde{x}_{\mbox{\scriptsize min}},\tilde{x}_{\mbox{\scriptsize max}}]$ and
$[\tilde{y}_{\mbox{\scriptsize min}},\tilde{y}_{\mbox{\scriptsize max}}]$,
through a point-to-point mapping from the original intervals into the new ones,
\begin{eqnarray*}
\left[x_{\mbox{\scriptsize min}},x_{\mbox{\scriptsize max}}\right] 
& \longrightarrow &
\left[\tilde{x}_{\mbox{\scriptsize min}},
      \tilde{x}_{\mbox{\scriptsize max}}\right], \quad \mbox{and} \\
\left[y_{\mbox{\scriptsize min}},y_{\mbox{\scriptsize max}}\right] 
& \longrightarrow &
\left[\tilde{y}_{\mbox{\scriptsize min}},
      \tilde{y}_{\mbox{\scriptsize max}}\right]
\end{eqnarray*}
For this purpose, linear transformations of the form
\begin{equation}
\tilde{x} = b_{x} x - c_{x} \quad \mbox{and} \quad
\tilde{y} = b_{y} y - c_{y}
\label{equation_linear_transformations}
\end{equation}
are appropriate, where $b$ and $c$ are constants ($b_x$ and $b_y$ are scaling
factors, and $c_x$ and $c_y$ represent origin offsets in the normalised data
ranges). The inverse transformations will be given by
\begin{equation}
x = \frac{\tilde{x}+c_x}{b_x} \quad \mbox{and} \quad
y = \frac{\tilde{y}+c_y}{b_y}.
\end{equation}
Assuming that the original and final intervals are not null (i.e.,
\mbox{$x_{\mbox{\scriptsize min}}\neq x_{\mbox{\scriptsize max}}$},
\mbox{$\tilde{x}_{\mbox{\scriptsize min}}\neq \tilde{x}_{\mbox{\scriptsize
max}}$}, \mbox{$y_{\mbox{\scriptsize min}}\neq y_{\mbox{\scriptsize max}}$} and
\mbox{$\tilde{y}_{\mbox{\scriptsize min}}\neq \tilde{y}_{\mbox{\scriptsize
max}}$}), it is trivial to show that the transformation constants are given by
\begin{eqnarray}
b_x & = & \displaystyle 
          \frac{\tilde{x}_{\mbox{\scriptsize max}}-
                \tilde{x}_{\mbox{\scriptsize min}}}
               {x_{\mbox{\scriptsize max}}-x_{\mbox{\scriptsize min}}},
\\ 
& & \nonumber \\ % extra empty line to separate both expressions
c_x & = & \displaystyle
          \frac{\tilde{x}_{\mbox{\scriptsize max}} x_{\mbox{\scriptsize min}}-
                \tilde{x}_{\mbox{\scriptsize min}} x_{\mbox{\scriptsize max}}}
               {x_{\mbox{\scriptsize max}}-x_{\mbox{\scriptsize min}}},
\end{eqnarray}
and the analogue expressions for the coefficients of the $y$-axis
transformation.  For example, to perform all the arithmetical manipulations
with small numbers, it is useful to choose 
\mbox{$\tilde{x}_{\mbox{\scriptsize min}}=
\tilde{y}_{\mbox{\scriptsize min}}\equiv -1$}
and
\mbox{$\tilde{x}_{\mbox{\scriptsize max}}=
\tilde{y}_{\mbox{\scriptsize max}}\equiv +1$}, 
which leads to
\begin{eqnarray}
b_x & = & \displaystyle 
          \frac{2}
               {x_{\mbox{\scriptsize max}}-x_{\mbox{\scriptsize min}}},
\label{equation_normalization_c_uno}
\\ 
& & \nonumber \\ % extra empty line to separate both expressions
c_x & = & \displaystyle
          \frac{x_{\mbox{\scriptsize min}}+x_{\mbox{\scriptsize max}}}
               {x_{\mbox{\scriptsize max}}-x_{\mbox{\scriptsize min}}},
\label{equation_normalization_c_dos}
\end{eqnarray}
and the analogue expressions for $b_y$ and $c_y$.

Once the data have been properly normalised in both axes following the
transformations given in Eq.~(\ref{equation_linear_transformations}), it is
possible to carry out the fitting procedure, which provides the
resulting polynomial expressed in terms of the transformed data ranges as
\begin{equation}
\tilde{y} = \tilde{a}_0+
            \tilde{a}_1 \tilde{x} +
            \tilde{a}_2 \tilde{x}^2 +
            \cdots +
            \tilde{a}_p \tilde{x}^p.
\end{equation}
At this point, the relevant question is how to transform the fitted
coefficients \mbox{$\tilde{a}_0,\tilde{a_1},\ldots,\tilde{a}_p$} into the
coefficients \mbox{$a_0,a_1,\ldots,a_p$} corresponding to the same polynomial
defined over the original data ranges.  By substituting the relations given in
Eq.~(\ref{equation_linear_transformations}) in the previous expression one
directly obtains
\begin{equation}
\begin{array}{@{}l@{\;}c@{\;}l}
(b_y y-c_y) & = & \tilde{a}_0+ \tilde{a}_1 (b_x x - c_x) +
                  \tilde{a}_2 (b_x x - c_x)^2 + \\
            &   & + \cdots + \tilde{a}_p (b_x x - c_x)^p.
\label{equation_replacing_transformations}
\end{array}
\end{equation}
Remembering that
\begin{equation}
(b_x x - c_x)^m = \displaystyle\sum_{n=0}^m 
                  \left( \begin{array}{@{}c@{}}m\\n\end{array} \right)
                  (b_x x)^{m-n} (-c_x)^n,
\label{equation_m_power}
\end{equation}
with the binomial coefficient computed as
\begin{equation}
\left( \begin{array}{@{}c@{}}m\\n\end{array} \right) = 
\frac{m!}{n! \; (m-n)!},
\label{equation_binomial_coefficient}
\end{equation}
and comparing the substitution of Eq.~(\ref{equation_m_power}) and
Eq.~(\ref{equation_binomial_coefficient}) into
Eq.~(\ref{equation_replacing_transformations}) with the expression given in
Eq.~(\ref{equation_fitted_polynomial}), it is not difficult to show that if
one defines
\begin{equation}
h_i \equiv \displaystyle\sum_{j=i}^{p} \tilde{a}_j
\left( \begin{array}{@{}c@{}}j\\j-i\end{array} \right)
(b_x)^i (-c_x)^{j-i}
\end{equation}
the sought coefficients will be given by
\begin{equation}
a_i = \left\{
\begin{array}{cl}
\displaystyle\frac{h_0+c_y}{b_y} & \mbox{for}\;i=0 \\
                                 &                 \\
\displaystyle\frac{h_i}{b_y}     & \mbox{with}\;i=1,\ldots,p
\end{array}
\right.
\label{equation:trans_pol}
\end{equation}
In the particular case in which $c_x=0$, the above expressions simplify to
\begin{equation}
a_i = \left\{
\begin{array}{cl}
\displaystyle\frac{\tilde{a}_0+c_y}{b_y} & \mbox{for}\;i=0 \\
                                 &                         \\
\displaystyle\frac{\tilde{a}_i b_x^i}{b_y}  & \mbox{with}\;i=1,\ldots,p
\end{array}
\right.
\label{equation:trans_pol_simple}
\end{equation}

\begin{figure}
\includegraphics[angle=0,width=\columnwidth]{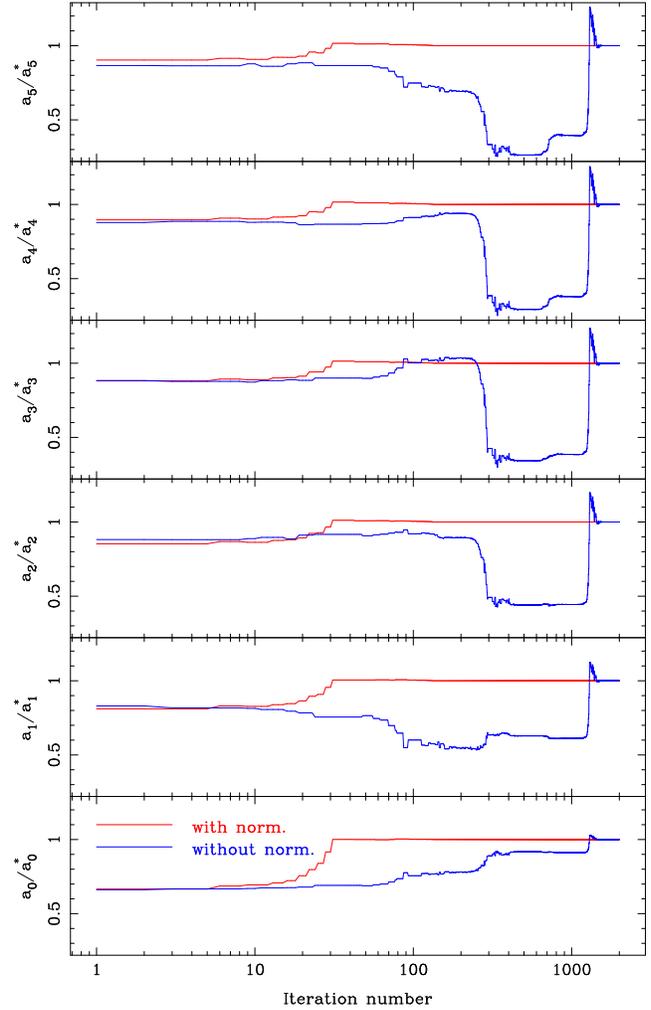}
\caption{Variation in the fitted coefficients, as a function of the number of
iterations, for the upper boundary fit (5th order polynomial) shown in
Fig.~\ref{figure:oneoverx_pol}a. This plot is the same than
Fig.~\ref{figure:coeff_pol}, but in this case analysing the impact of the
normalisation of the data ranges prior to the boundary determination. Each
panel represents the coefficient value at a given iteration ($a_i$, with
\mbox{$i=0,\ldots,5$}, from bottom to top) divided by $a_i^{*}$, the final
value derived after \mbox{$N_{\mbox{\scriptsize maxiter}}=2000$} iterations.
The same \mbox{$y$-axis} range is employed in all the plots.
The red line shows the results when applying the normalisation, and the blue
line indicates the coefficient variations when this normalisation is not
applied. In both cases \mbox{$\xi=1000$}, \mbox{$\alpha=2$} and
\mbox{$\beta=0$} were used. Note that the plot \mbox{$x$-scale} is in
logarithmic units.}
\label{figure:coeffnorm}
\end{figure}

The normalisation of the data ranges has several advantages.
Fig.~\ref{figure:coeffnorm} (similar to Fig.~\ref{figure:coeff_pol}) shows
the impact of data normalisation on the convergence properties of the fitted
coefficients, as a function of the number of iterations, for the upper boundary
fit (5th order polynomial) shown in Fig.~\ref{figure:oneoverx_pol}a. The
red line, corresponding to the results when the normalisation is applied
prior to the boundary fitting, indicates that after \mbox{$N_{\mbox{\scriptsize
maxiter}}\sim 140$}, the coefficients have converged. The situation is much
worse when the normalisation is not applied, as illustrated by the blue line.
In this case the convergence is only reached after 
\mbox{$N_{\mbox{\scriptsize maxiter}}\sim 1450$} iterations, ten times more
than when using the normalisation. In addition, the ranges spanned by the
coefficient values along the minimisation procedure are narrower when the data
ranges have been previously normalised.

\begin{figure}
\includegraphics[angle=0,width=\columnwidth]{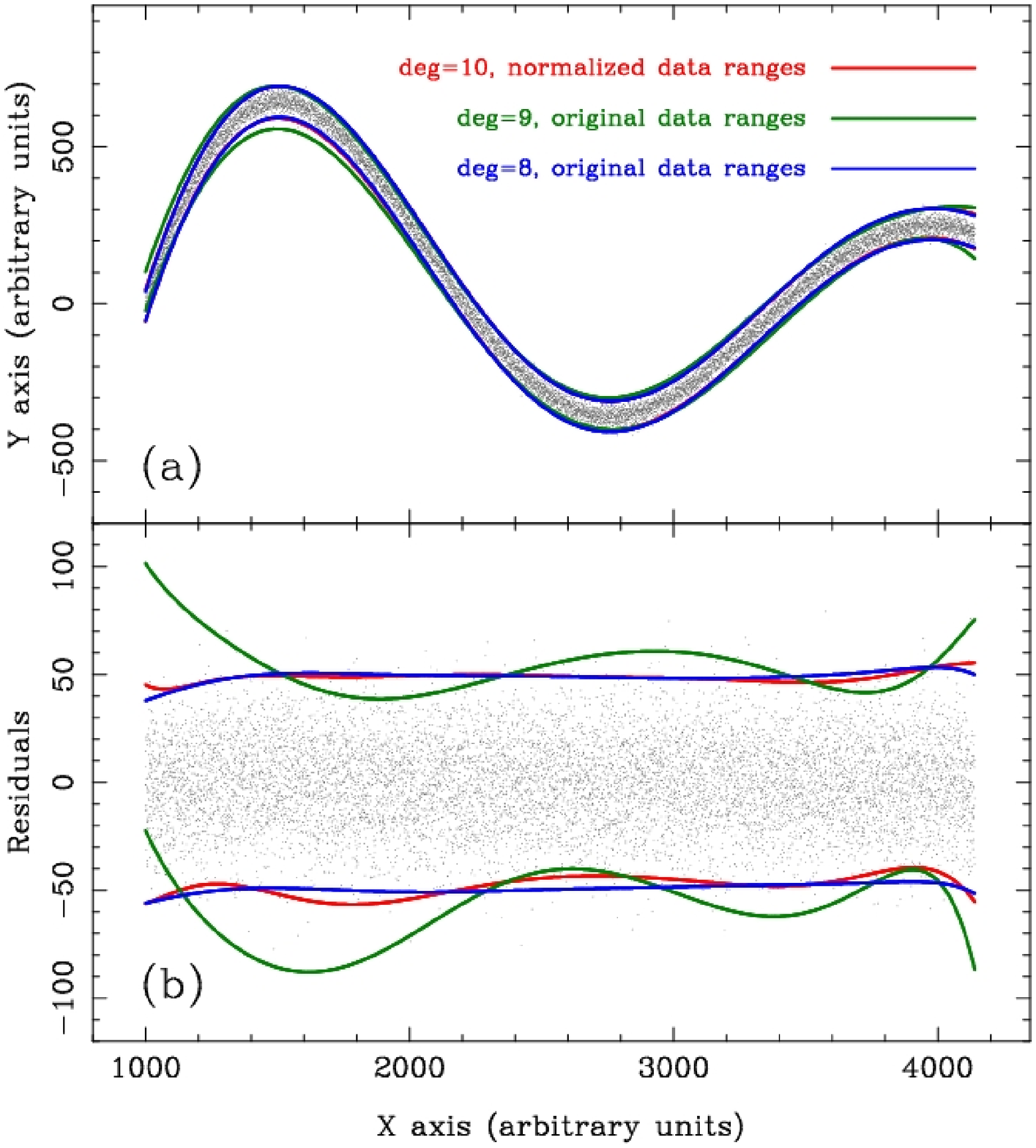} 
\caption{Example of the appearance of numerical errors in the boundary fitting
with simple polynomials.  The fitted data set consists in 10000 points randomly
drawn from the function \mbox{$y=\sin(1.5 x)/(1+x)$} for \mbox{$x\in[0,2\pi]$},
assuming a Gaussian error \mbox{$\sigma=0.02$} in the \mbox{$y$-axis}, and
where prior to the data fitting the $(x,y)$ coordinates were transformed using
\mbox{$x_{\mbox{\scriptsize fit}}=1000+500\; x_{\mbox{\scriptsize original}}$}
and \mbox{$y_{\mbox{\scriptsize fit}}=1000\; y_{\mbox{\scriptsize original}}$}
in order to artificially enlarge the data ranges.  \emph{Panel~(a)}:
bootstrapped data and fitted boundaries. \emph{Panel~(b)}: residuals relative
to the original sinusoidal function. In both panels the lines indicate the
resulting fits for different polynomial degrees and normalisation strategies
(in all the cases \mbox{$\xi=1000$}, \mbox{$\alpha=2$} and \mbox{$\beta=0$}
were employed). The continuous red lines are the boundaries obtained using
polynomials of degree 10 and normalising the data ranges prior to the fitting
procedure. The green and blue lines correspond to the fits obtained by fitting
polynomials of degrees 9 and 8, respectively, without normalising the data
ranges. Using the original data ranges the boundary fits start to depart from
the expected location due to numerical errors for polynomials of degree 9.
However polynomials of degree 10 are still an option when the data ranges are
previously normalised.}
\label{figure:numerr}
\end{figure}

Fig.~\ref{figure:numerr} exemplifies the appearance of numerical errors that
takes place when increasing the polynomial degree during the fitting of a
reasonably large data set. In this case 10000 points are fitted employing upper
and lower boundaries with simple polynomials of degree 10 (red lines) after
normalising the data ranges using the coefficients given in
Eqs.~(\ref{equation_normalization_c_uno})
and~(\ref{equation_normalization_c_dos}) (with the analogue expressions for the
$y$-axis coefficients) prior to the numerical minimisation. When the data
ranges are not normalised, the fitting to polynomials of degree 10 gives
non-sense results. Only polynomials of degree less or equal than 9 are
computable. And for the case of degree 9 the results are unsatisfactory (green
lines), being the polynomials of degree 8 (blue lines) the first reasonable
boundaries while fitting the data preserving their original ranges. Thus in
this particular example the normalisation of the data ranges allows to extend
the fitted polynomial degree in two units.

%------------------------------------------------------------------------------
\subsection{Cubic splines}

Normalisation of the data ranges is also important for the computation of cubic
splines, in particular for the boundary fitting to adaptive splines described
in Section~\ref{section:adaptive_splines}. In that section the functional form
of a fit to set of $N_{\mbox{\scriptsize knots}}$ was expressed as
\begin{equation}
\begin{array}{@{}r@{\;}l}
y     = & s_3(k) [x-x_{\mbox{\scriptsize knot}}(k)]^3 +
          s_2(k) [x-x_{\mbox{\scriptsize knot}}(k)]^2 + \nonumber \\
      + & s_1(k) [x-x_{\mbox{\scriptsize knot}}(k)] + s_0(k),
\end{array}
\label{equation:splines_original}
\end{equation}
where ($x_{\mbox{\scriptsize knot}}(k),y_{\mbox{\scriptsize knot}}(k)$) are the
$(x,y)$ coordinates of the $k^{\mbox{\scriptsize th}}$~knot, and $s_0(k)$,
$s_1(k)$, $s_2(k)$, and $s_3(k)$ are the corresponding spline coefficients for
\mbox{$x\in[x_{\mbox{\scriptsize knot}}(k),x_{\mbox{\scriptsize knot}}(k+1)]$},
with \mbox{$k=1,\ldots,N_{\mbox{\scriptsize knots}}-1$}.

Using the same nomenclature previously employed for the case of simple
polynomials, the result of a fit to cubic splines performed over normalised
data ranges should be written as
\begin{equation}
\begin{array}{@{}r@{\;}l}
\tilde{y}  = & 
      \tilde{s}_3(k) [\tilde{x}-\tilde{x}_{\mbox{\scriptsize knot}}(k)]^3 +
      \tilde{s}_2(k) [\tilde{x}-\tilde{x}_{\mbox{\scriptsize knot}}(k)]^2 + 
      \nonumber \\
      + & \tilde{s}_1(k) [\tilde{x}-\tilde{x}_{\mbox{\scriptsize knot}}(k)] + 
      \tilde{s}_0(k).
\end{array}
\label{equation:splines_normalized}
\end{equation}
%The relevant question, again, is how to compute the coefficients $s_i$ in
%Eq.~(\ref{equation:splines_original}) from $\tilde{s}_i$ in
%Eq.~(\ref{equation:splines_normalized}), with \mbox{$i=0,\ldots,3$}. To answer
%this question one only has to substitute the transformations given in
%Eqs.~(\ref{equation_normalization_c_uno})
%and~(\ref{equation_normalization_c_dos}) (and the corresponding normalisation
%for the $y$-axis) into
%Eq.~(\ref{equation:splines_normalized}) and operate. In particular
%\begin{equation}
%\begin{array}{@{}r@{\;}l}
%(b_y y - c_y) = & 
%     \tilde{s}_3(k) [(b_x x -c_x)-(b_x x_{\mbox{\scriptsize knot}}(k)-c_x)]^3 +
%     \nonumber \\
% + & \tilde{s}_2(k) [(b_x x -c_x)-(b_x x_{\mbox{\scriptsize knot}}(k)-c_x)]^2 + 
%     \nonumber \\
% + & \tilde{s}_1(k) [(b_x x-c_x)-(b_x x_{\mbox{\scriptsize knot}}(k)-c_x)] + 
%     \nonumber \\
% + & \tilde{s}_0(k).
%\end{array}
%\end{equation}
%The previous expression immediately simplifies to
%\begin{equation}
%\begin{array}{@{}r@{\;}l}
%(b_y y - c_y) = & 
%     \tilde{s}_3(k) b_x^3 [x -x_{\mbox{\scriptsize knot}}(k)]^3 +
%     \nonumber \\
% + & \tilde{s}_2(k) b_x^2 [x -x_{\mbox{\scriptsize knot}}(k)]^2 + 
%     \nonumber \\
% + & \tilde{s}_1(k) b_x   [x- x_{\mbox{\scriptsize knot}}(k)] + 
%     \nonumber \\
% + & \tilde{s}_0(k).
%\end{array}
%\end{equation}
%Comparing this result with Eq.~(\ref{equation:splines_original})
%it is straightforward to see that the sought transformations are
Following a similar reasoning to that used previously, it is
straightforward to see that the sought transformations are
\begin{equation}
s_i(k) = \left\{
\begin{array}{cl}
\displaystyle\frac{\tilde{s}_0(k)+c_y}{b_y} & \mbox{for}\;i=0 \\
                                 &                         \\
\displaystyle\frac{\tilde{s}_i(k) b_x^i}{b_y} & \mbox{with}\;i=1,\ldots,3
\end{array}
\right.
\end{equation}
where \mbox{$k=1,\ldots,N_{\mbox{\scriptsize knots}}-1$}. Note that these
transformations are identical to Eq.~(\ref{equation:trans_pol_simple}). This is
not surprising considering that splines are polynomials and that the adopted
functional form given in Eq.~(\ref{equation:splines_original}) is actually
providing the $y(x)$ coordinate as a function of the distance between the
considered $x$ and corresponding value $x_{\mbox{\scriptsize knot}}(k)$ for the
nearest knot placed at the left side of $x$. Thus, the $c_x$ coefficient is not
relevant here.

%------------------------------------------------------------------------------
\subsection{A word of caution}

Although the method described in this appendix can help in some circumstances
to perform fits with larger data sets or higher polynomial degrees than without
any normalisation of the data ranges, it is important to keep in mind that such
normalisation does not always produce the expected results and that numerical
errors appear in any case sooner or later if one tries to use excessively large
data sets or very high values for the polynomial degrees. 

Anyhow, the fact that the normalisation of the data ranges can facilitate the
boundary determination of large data sets or to use higher polynomial degrees
justifies the effort of checking whether such normalisation is of any help.
Sometimes, to extend the polynomial degrees by even just a few units can be
enough to solve the particular problem one is dealing with. The program {\tt
BoundFit} incorporates the normalisation of the data prior to the boundary
fitting as an option.

%%%%%%%%%%%%%%%%%%%%%%%%%%%%%%%%%%%%%%%%%%%%%%%%%%%%%%%%%%%%%%%%%%%%%%%%%%%%%%%
%%%%%%%%%%%%%%%%%%%%%%%%%%%%%%%%%%%%%%%%%%%%%%%%%%%%%%%%%%%%%%%%%%%%%%%%%%%%%%%
\label{lastpage}

\end{document}